\documentclass[twocolumn,english,aps,jcp,superscriptaddress,showpacs,floatfix]{revtex4-2} 
\usepackage{graphicx}
\usepackage{subfigure}
\usepackage{amsmath,amssymb,amsfonts}
\usepackage{changes}
\usepackage{babel}
\usepackage{tikz}
\usepackage{xcolor}
\usepackage[shortlabels]{enumitem}
\usetikzlibrary{shapes,arrows}
\usepackage{dcolumn}
\usepackage{bm}

\DeclareMathAlphabet\mathbfcal{OMS}{cmsy}{b}{n}

\begin{document}

\title{Collective photon emission in solid state environments: Concatenating non-markovian and markovian dynamics}%

\author{Devashish Pandey}%
\affiliation{Department of Electrical and Photonics Engineering, Technical University of Denmark, Kgs. Lyngby, Denmark}
\author{Martijn Wubs}%
\email{mwubs@dtu.dk}
\affiliation{Department of Electrical and Photonics Engineering, Technical University of Denmark, Kgs. Lyngby, Denmark}
\affiliation{NanoPhoton - Center for Nanophotonics,\\ Technical University of Denmark, Kgs. Lyngby, Denmark}

\begin{abstract}
Collective light emission and multi-qubit dynamics of solid-state quantum emitters are affected both by their coupling to the light field and to lattice vibrations. The effect of phonons on quantum emitters is twofold: polaron formation is described by ultrafast non-markovian dynamics, while slower dephasing is well described by exponential decay.  Of the two temperature-dependent processes, the effect of the former on the collective emission and the entanglement decay of emitters is usually not modeled, and also the latter is sometimes neglected. Here we propose and compare two methods that are efficient also for several emitters: the first method concatenates the fast and slow phonon dynamics, and the second is the polaron method. For a single quantum emitter, we show that the dynamical equations are identical in both methods, while predictions for two or more emitters also agree very well. Both of our methods incorporate non-markovian dynamics due to phonons demonstrating the temperature sensitivity of the collective photon emission. Utilizing a simplified markovian model instead may not be accurate enough especially for quantum information applications: for example, we show how the markovian model may considerably overestimate the two-emitter concurrence, except at very low temperatures. Our concatenation and polaron methods can be applied to an arbitrary number and type of quantum emitters, and beyond the bulk GaAs environment that we consider here. Especially the concatenation method can take phonon effects into account at the same computational cost as modelling the emitter-photon interaction alone. Finally, we present approximate analytical expressions for the collective emission spectrum for $N$ emitters on a one-dimensional chain. 
\end{abstract}
\maketitle

\section{Introduction} 
\label{sec:intro}
With quantum information processing and sensing applications in mind, in recent years, there has been a surge of research in generating single-photon emitters. Various strategies have been explored to fabricate arrays of them~\cite{Lagoin2023Dual-densityBilayers,Toninelli2021SingleTechnologies,Parto2021DefectK,Fournier2021Position-controlledNitride}, which may exhibit  collective emission (sub- and superradiance) with potential applications in photovoltaics~\cite{Fruchtman2016PhotocellProtection,Rouse2019OptimalEnvironment}, energy harvesting complexes~\cite{Tomasi2021EnvironmentallyHarvesting, Moreno-Cardoner2022EfficientRings}, superradiant lasers \cite{Bohnet2012APhoton,Protsenko2021QuantumNanolasers}, and quantum memories \cite{Lvovsky2009OpticalMemory}. Superradiance at very low temperatures has already been demonstrated in solid-state systems such as quantum dots and nitrogen-vacancy centers in diamond~\cite{Scheibner2007SuperradianceDots, Cong2016DickeInvited, Angerer2018SuperradiantDiamond}. Phonon-induced decoherence make the observation of room-temperature superradiance usually very challenging, so its recent observation in perovskites is remarkable~\cite{Biliroglu2022Room-temperatureOrigins}. Another recent breakthrough is the observation of controllable collective emission by two distant quantum dots~\cite{Tiranov2023CollectiveEmitters}. \par
Considerable work has already been dedicated to modeling the collective effects with emitters interacting with only the photon bath~\cite{Lehmberg1970RadiationFormalism, Gruner1996Green-functionDielectrics, Wubs2004Multiple-scatteringDielectrics, ficek2005quantum, Asenjo-Garcia2017ExponentialArrays, Zhang2020SubradiantTheorem, Feist2020, Reitz2022CooperativePlatforms, Sanchez-Barquilla2022Few-modeEmitters, Moreno-Cardoner2022EfficientRings, Jrgensen2022QuantifyingSuperradiance}. 
Usually, the markovian approximation is excellent for the photon bath, as spectral features of the bath do not vary much within the linewidths of the emitters. In the strong-coupling regime of cavity QED and in some other carefully engineered situations~\cite{Vats1998Non-MarkovianEdge, Nikolaev2007StronglyCrystals, Madsen2011ObservationCavity} the markovian approximation for the photon bath is no longer valid. Such situations are not considered here.\par
On the other hand, solid-state emitters also interact with phonon baths, for which the markovian approximation is typically less well justified~\cite{Rouse2019OptimalEnvironment, Tomasi2021EnvironmentallyHarvesting, Holzinger2022CooperativeArrays, Koong2022CoherenceEmitters, Wiercinski2023PhononEmitters, PhysRevX.14.011010}. One of the main goals of our work is to propose novel and computationally efficient methods for collective emission in solid-state environments, to obtain simple analytical results that give maximum physical insight with few approximations.\par
Already for a single solid-state emitter, modeling the emission characteristics can become a complex affair, as it depends on both emitter-photon and emitter-phonon interactions, but master equations that account for both types of interactions have already been utilized to explain and stimulate fascinating experiments~\cite{Ramsay2010DampingDots, McCutcheon2010QuantumRegime, Hughes2013Phonon-mediatedSystem, Quilter2015Phonon-AssistedExcitation}. A weak system-bath coupling justifies the common second-order perturbation approximation, resulting in computationally convenient master equations of the Redfield or Lindblad types. These equations become less accurate not only under stronger system-bath coupling but also at higher temperatures, as they do not capture the non-markovian dynamics of the quantum emitter due to interactions with phonons~\cite{Ramsay2010DampingDots, McCutcheon2010QuantumRegime,Nazir2016ModellingDots}.\par
The non-markovian dephasing dynamics of a quantum emitter in a bulk phonon bath is typically orders of magnitude faster than its photon emission dynamics. This can be exploited in calculations of the dynamics to obtain both physical insight and computational efficiency, for example, in the initial-slip scheme and its augmented form the concatenation scheme~\cite{Geigenmuller1983SystematicSystems, Haake1985StrongOscillator, Suarez1992MemorySystems, Yu2000Post-markovSystems, Cheng2005MarkovianSystems}. In the initial-slip scheme, the effect of the ultrafast initial non-markovian phonon dynamics is accounted for only by modifying the initial state for a markovian master equation of the quantum system~\cite{Suarez1992MemorySystems, Yu2000Post-markovSystems}. The concatenation scheme, on the other hand, concatenates the initial non-markovian dynamics obtained from the second-order perturbation theory with the subsequent markovian dynamics, giving a more accurate description of the system dynamics at all relevant time scales~\cite{Cheng2005MarkovianSystems}.\par
Here we adopt such a concatenation approach, but we generalize it in three important ways: first, we replace the second-order perturbation theory with the exact fast first-order phonon dynamics. Second, we also take the slower second-order phonon dynamics into account, in the markovian approximation, following Refs.~\cite{Muljarov2004DephasingPhonons, Reigue2017ProbingDots, Tighineanu2018PhononDimensionality}. Third, we make a generalization from one emitter to $N$ interacting emitters.\par
We compare multi-emitter spectra in the resulting concatenation approach with spectra obtained with a second powerful approach to deal with non-markovian phonon dynamics, namely the well-known polaron approach 
~\cite{Duke1965Phonon-broadenedStates, Silbey1983VariationalBath, Leggett1987DynamicsSystem, Mahan2000Many-ParticlePhysics,Wubs2006GaugingTransitions, Nazir2009Correlation-dependentDynamics, McCutcheon2010QuantumRegime, McCutcheon2011CoherentEffects, Nazir2016ModellingDots, Reigue2017ProbingDots, Iles-Smith2017LimitsEmitters, Maguire2019EnvironmentalSystems, Rouse2019OptimalEnvironment, Denning2020PhononSources, Cosacchi2021AccuracyDot, Tomasi2021EnvironmentallyHarvesting, Rouse2022AnalyticFrame, Reigue2017ProbingDots, Brash2019LightPicture, Koong2019FundamentalTransitions, Brash2023NanocavityK, Fischer2023CombiningOrigin}, that we here also generalize to $N$ emitters.

The polaron master equation is obtained after transforming the Hamiltonian 
into the so-called polaron frame, resulting in non-additive interactions between photon and phonon baths~\cite{Maguire2019EnvironmentalSystems,Rouse2022AnalyticFrame}. 
The polaron approach is the more accurate yet more complex approach of the two. Therefore, one of our main aims is to study how well the multi-emitter spectra agree in the two approaches. 

Although we will not make use of them here, it is important to mention that there also exist numerically exact methods~\cite{Vagov2007NonmonotonicDots,Barth2016Path-integralSystems, Richter2022EnhancedNetworks, Cygorek2022SimulationEnvironments}. We study collective emission under the influence of phonons as an initial-value problem. By not explicitly modeling the excitation pulse, we work in the weak-driving limit, where the polaron method becomes exact~\cite{Cosacchi2021AccuracyDot}. For strong driving in combination with strong phonon coupling, not considered here, the exact numerical approaches are better suited~\cite{Vagov2007NonmonotonicDots,Barth2016Path-integralSystems, Richter2022EnhancedNetworks, Cygorek2022SimulationEnvironments}. An advantage of using  the polaron method for our analysis is that it allows us to obtain analytical results that can be compared with the concatenation approach. Moreover, both methods can be generalized to  multi-emitter problems in computationally efficient ways.\par

In Sec.~\ref{sec:2}, we describe the interaction of the emitter(s) with the phonon and photon baths using the concatenation scheme and the polaron method for three different scenarios corresponding to one, two and more than two emitters. First, in Sec.~\ref{sec:1e} we show that the agreement between our non-markovian concatenation and polaron methods for single emitter stems from different timescales of emitter-phonon and emitter-photon interactions. In Sec.~\ref{sec:2e}, we extend this to two emitters, where the agreement between the two methods is no longer exact, because only the polaron method accounts for the non-additive interaction between the phonon and photon bath, resulting in phonon-bath dependent collective decay rates. Still our quantitative predictions from both methods agree very well. In this section we also discuss the entanglement dynamics between two emitters (``qubits”) which is overestimated by the commonly used markovian models. Our non-markovian models on the other hand give analytical predictions for their fast initial entanglement decay at finite temperatures. We expect collective emission by three or more solid-state emitters to become experimentally feasible in the near future. Thus, in Sec.~\ref{sec:Ne}, we generalize the polaron method and our concatenation scheme for $N$ emitters in a 1D chain. Both methods can efficiently describe non-markovian effects on collective emission by many emitters. As an example we will present results for $N = 8$. We conclude in Sec.~\ref{sec:conclusions}. Furthermore, we support our main results by detailed calculations in the Appendices.
\section{Emitter(s) interacting with photon and phonon baths}
\label{sec:2}
\subsection{Scenario I : $N = 1$}
\label{sec:1e}
Let us now introduce our model Hamiltonian and some assumptions, first for a single emitter, before discussing single-emitter and multi-emitter spectra.
An emitter coupled to both the phonon and photon bath can be described by the Hamiltonian
\begin{equation}
H^{\rm 1e} = H_0 + H_{\rm e-pt} + H_{({\rm e-pn})_1} + H_{({\rm e-pn})_2} + H_{\rm pt} + H_{\rm pn},
\label{H1e}
\end{equation}
where the superscript `1e' indicates that here we are first considering the Hamiltonian of a single emitter. Here $H_{\rm pt} = \sum_q\omega_qa_q^\dagger a_q$  and $H_{\rm pn} =  \sum_{k}\omega_{k} b_{k}^\dagger b_{k}$ are the photon and phonon bath Hamiltonians, respectively and we work in the units of $\hbar = 1$. $H_0 = \omega_0\sigma^\dagger\sigma$ is the system Hamiltonian describing a two-level emitter with resonance frequency $\omega_0$, while $\sigma$, ($\sigma^\dagger$) is the lowering (raising) atomic operator.  $H_{\rm e-pt} = \sum_q(h_q a_q + h_q ^*a_q^\dagger)\sigma^\dagger + \mbox{h.c.}$, describes the emitter-photon coupling modeled within the dipole approximation, with $h_q$ the coupling between the emitter and the photon field of mode $q$ and $a_q$ ($a_q^\dagger$) is the corresponding photon annihilation (creation) operator. The mode index $q$ labels both the wavevector and corresponding polarization indices of transverse plane waves. Likewise, the $b_k$ ($b_k^\dagger$) are phonon annihilation (creation) operators.\par
To be specific, we will use the material parameters that describe the emitter-acoustic phonon coupling in bulk GaAs~\cite{Blakemore1982SemiconductingArsenide, Cardona1987AcousticSemiconductors, Tighineanu2018PhononDimensionality,  Denning2020PhononSources}. The dominant electron-phonon coupling in quantum dots is deformation-potential coupling to longitudinal acoustic (LA) phonons, and we will disregard other couplings~\cite{Krummheuer2002TheoryDots}.
This results in a three-dimensional (3D) phonon spectral density $J_{\rm pn}(\omega) = \sum_k |g_k|^2\delta(\omega - \omega_k) = \alpha \omega^3\mbox{exp}(-\omega^2/\omega_{\rm c}^2)$ where $g_k$ is the emitter-phonon coupling constant, $\alpha$ is the deformation potential coupling constant and $\omega_{\rm c}$ is the cutoff frequency which is given by the ratio of the speed of sound and the size of the QD~\cite{Nazir2016ModellingDots}. For the light-matter coupling, for simplicity we will also consider bulk GaAs. We neglect photon loss, which is an excellent approximation at inter-emitter distance scales of micrometers or below on which emission can be collective.\par
Our model Hamiltonian defined in \eqref{H1e} contains two types of emitter-phonon interactions that both will affect emission spectra. In the presence of phonons, an excited emitter undergoes a fast non-markovian dephasing on the picosecond scale, followed by a slower dephasing responsible for the temperature-dependent broadening of the zero-phonon line (ZPL)~\cite{Borri2001UltralongDots, Muljarov2004DephasingPhonons, Reigue2017ProbingDots} (see Appendix~A).  The linear emitter-phonon coupling $H_{{(\rm e-pn)}_1} =  \sigma^\dagger\sigma\sum_k g_k b_{k}^\dagger + \mbox{h.c.}$ is responsible for the fast dephasing.\par
The excitation of the emitter leads to a sudden change in the charge configuration of the lattice, resulting in a fast decay of coherence of the emitter until the lattice energy is renormalized to a new equilibrium. At this point, a polaron is formed in the ionic lattice, at the characteristic time $\tau= \tau_{\rm P}$~\cite{Hohenester2007QuantumDots}. For GaAs, we estimate $\tau_{\rm P} \approx 2$ ps (see Appendix~A). The slower dephasing is modeled by a quadratic coupling between the emitter and phonons $H_{{(\rm e-pn)}_2} = \sigma^\dagger\sigma\sum_{k,k'} f_{k,k'} (b_{k}^\dagger + b_k)(b_{k'}^\dagger + b_{k'})$ where $f_{k,k'}$ describes phonon-assisted virtual transitions to higher-lying states of the emitter that cause an inelastic scattering of phonons from mode $k$ to $k'$ \cite{Muljarov2004DephasingPhonons, Reigue2017ProbingDots}. This leads to markovian pure dephasing of the excited state of the emitter with a dephasing time $1/\gamma_{\rm pd}$, where $\gamma_{\rm pd}$ is the dephasing rate~\cite{Muljarov2004DephasingPhonons}.\par
Having specified our complete Hamiltonian, we now first calculate single-emitter dynamics in the concatenation approach, which is expected to be accurate because the initial polaron formation is much faster than spontaneous emission and markovian phonon pure dephasing. The initial non-markovian dynamics is therefore governed by the Hamiltonian $H_{\rm NM}^{\rm 1e} = H_0 + H_{{(\rm e-pn)_1}}  + H_{\rm pn}$, which is an instance of the exactly solvable independent boson model (see Refs.~\cite{Duke1965Phonon-broadenedStates,Mahan2000Many-ParticlePhysics,Krummheuer2002TheoryDots} and Appendix~B). We then use the exact unitary operator associated with $H_{\rm NM}^{\rm 1e}$, to calculate the two-time field correlation~\cite{ficek2005quantum} using the non-markovian evolution (see~\cite{Nazir2016ModellingDots} and Appendix~B),
\begin{equation}
  g^{(1)}(\tau)=  \gamma\langle \sigma^\dagger\sigma(\tau)\rangle_{\rm NME}=
    \gamma\rho_{\rm ee} \mathcal{C}(\tau)e^{-i\omega_0'\tau}, \;\; \tau \leq \tau_{\rm P}. 
\label{corr_nonmark_1e}
\end{equation}
Here $\rho_{\rm ee} = \langle e | \rho(0) |e  \rangle $ is the population of the excited state at the initial time $t = 0$,  $\mathcal{C}(\tau) =\mathcal{C}_{\infty}(T) \mbox{exp}(\phi(\tau))$ is the phonon correlation function with $\mathcal{C}_\infty(T) = \mbox{exp}(-\int_0^\infty d\omega J_{\rm pn}(\omega) \coth(\beta\omega/2)/\omega^2)$ its steady-state value, where $\beta = 1/k_{\rm B}T$ is the inverse temperature and $k_{B\rm }$ the Boltzmann constant. Notice that we have made the temperature dependence of this steady-state value explicit, which we will use henceforth. $\mathcal{C}_\infty(T)$ is referred to as the Franck-Condon factor and describes the overlap of the displaced phonon state and the phonon ground state. Furthermore, $\phi(\tau) = \int_0^\infty d\omega J_{\rm pn}(\omega)\left(\cos(\omega\tau)\coth(\beta\omega/2)-i\sin(\omega\tau))\right/\omega^2$ is the phonon propagator and  $\omega_0' = \omega_0 - \int d\omega J_{\rm pn}(\omega)/\omega$,  where the reorganization energy $\int d\omega J_{\rm pn}(\omega)/\omega$ is also known as the polaron shift~\cite{Mahan2000Many-ParticlePhysics}.\par
The subsequent dynamics at $\tau > \tau_{\rm P}$ is modeled within the markovian approximation and encompasses both spontaneous decay and the quadratic emitter-phonon coupling, as described by the Hamiltonian $H_{\rm MK}^{\rm 1e} = H_0 + H_{\rm e-pt} + H_{{(\rm e-pn)}_2} + H_{\rm pt} + H_{\rm pn}$. This leads in the usual way to a  master equation of the Lindblad form, ${d\rho}/{dt}= \gamma \mathcal{D}_{\sigma}(\rho) + 2\gamma_{\rm pd}(T)\mathcal{D}_{\sigma^\dagger\sigma}(\rho)$, where $\mathcal{D}_{\sigma}(\rho) =(\sigma\rho\sigma^\dagger-1/2\{\sigma^\dagger\sigma,\rho\}_+)$ describes the spontaneous decay with a decay rate $\gamma$, and where $\{\}_{+}$ is the anticommutator.  The dissipator $\mathcal{D}_{\sigma^\dagger\sigma}(\rho)$ describes the temperature-dependent pure dephasing of the excited emitter state at the rate $\gamma_{\rm pd}(T)$  due to the quadratic phonon interaction, which has been calculated for GaAs quantum dots in~\cite{Reigue2017ProbingDots, Tighineanu2018PhononDimensionality} (see also Appendix~A). Our concatenation approach is consistent as long as markovian phonon pure dephasing and spontaneous decay are negligible on the time scale $\tau_{\rm P}$, i.e. $\mbox{exp}(-\Gamma \tau_{\rm P})\approx 1$, where $\Gamma \equiv {\gamma}/{2} + \gamma_{\rm pd}(T)$. For times $\tau > \tau_{\rm P}$ we can apply the quantum regression theorem (QRT)~\cite{Carmichael2000StatisticalEquations} and find from our concatenation approach at all times the field correlation (see Appendix~B for details) 
\begin{equation}
    g^{(1)}(\tau) = \gamma\rho_{\rm ee}\mathcal{C}(\tau)\mbox{e}^{-(\Gamma + i\omega_0')\tau}.\;\;\;
     \label{g_1_1e}
\end{equation}
This first-order correlation function can be measured experimentally~\cite{Borri2001UltralongDots, Brash2019LightPicture}.\par
Interestingly, for a single quantum emitter, polaron theory predicts exactly the same two-time correlation as in~\eqref{g_1_1e}~\cite{Chassagneux2018EffectRegime, Iles-Smith2017LimitsEmitters, Reigue2017ProbingDots, Nazir2016ModellingDots, Iles-Smith2017PhononSources}. We will introduce the polaron theory below in Sec.~\ref{sec:2e}, but in brief, the equivalence between both approaches is related to the fact that also in polaron theory, it is assumed that the polaron formation happens on a faster time scale than all other processes. We think it is simpler and offers more physical insight to use our concatenation approach to arrive at Eq.~(\ref{g_1_1e}). Regardless, both methods agree on markovian and non-markovian effects due to the emitter-phonon coupling.\par
For later use, we also briefly review the corresponding known single-emitter frequency spectrum that can be obtained from the first-order correlation function by utilizing the optical Wiener-Khinchin theorem~\cite{Carmichael2000StatisticalEquations}. In the frequency spectrum $S(\omega)$, the fast initial phonon-assisted non-markovian decay of the correlations results in broad meV-range sidebands, while the subsequent slow markovian decay due to photons and phonons results in a finite linewidth of the ZPL due to spontaneous emission and a further broadening of the ZPL  due to pure phonon dephasing~\cite{Duke1965Phonon-broadenedStates, Mahan2000Many-ParticlePhysics, Muljarov2004DephasingPhonons, Reigue2017ProbingDots, Denning2020PhononSources}. In  Appendix~A we depict the emission spectrum using~\eqref{g_1_1e}, demonstrating the temperature dependence of photon emission into the ZPL and sidebands.\par
The total spectrum $S(\omega)$ can be written as $S_{\rm ZPL}(\omega) + S_{\rm SB}(\omega)$, i.e. the sum of the contributions from the ZPL and from the sidebands. 
The contribution of only the ZPL to the total spectrum can be singled out by making in Eq.~(\ref{g_1_1e}) the replacement  $\mathcal{C}(\tau)\rightarrow\mathcal{C}_\infty(T)$~\cite{Iles-Smith2017LimitsEmitters, Rouse2019OptimalEnvironment}. In this approximation, the concatenation scheme reduces to the initial-slip scheme where $\mathcal{C}_\infty(T)$ provides the initial condition for the evolution of the markovian master equation.  
The approximation results in an analytical expression for the spectrum near the zero-phonon line of the form $S_{\rm ZPL}^{\rm 1e}(\omega) = \mathcal{C}_{\infty}(T)\gamma\rho_{\rm ee}\Gamma/((\omega - \omega_0')^2 + \Gamma^2)$. Thus $\mathcal{C}_\infty(T)$ also measures the proportion of the emission that is channeled into the ZPL~\cite{Iles-Smith2017LimitsEmitters, Iles-Smith2017PhononSources}. Materials and operating temperatures $T$ such that  $\mathcal{C}_\infty(T)$ is closer to unity are more interesting for quantum information processing applications. To summarize this section, we studied the dynamics of a single emitter interacting with both phonon and photon baths, and we found an exact agreement between our concatenation and polaron methods.
\subsection{Scenario II : $N = 2$}
\label{sec:2e}
Next, we extend our approach to two identical emitters. Frequency tuning real solid-state emitters such that they become optically identical is an art in itself~\cite{Koong2022CoherenceEmitters,Tiranov2023CollectiveEmitters}. For definiteness, we assume the two emitters to be embedded in bulk GaAs and separated by a distance $r_{12}$, both interacting with the photon and phonon baths. This can be described by the Hamiltonian
\begin{multline}
H^{\rm 2e} = \sum_{n=1}^2 \left(H_0^{(n)} + H_{\rm e-pt}^{(n)} + H_{({\rm e-pn})_1}^{(n)} + H_{({\rm e-pn})_2}^{(n)} \right)\\ + H_{\rm pt} + H_{\rm pn},
\label{H2e}
\end{multline}
where we use the same nomenclature as in Sec.~\ref{sec:1e} except that now the coupling constants $h_q^{(n)}$, $g_k^{(n)}$ and $f_{k,k'}^{(n)}$ depend on the position of the $n$-th emitter.\par
To solve the two-emitter dynamics, we first use our concatenation method. From the Heisenberg equations of motion corresponding to the Hamiltonian~\eqref{H2e}, we find the first-order correlation function $g^{(1)}(\tau) = \sum_{n,m=1}^2\gamma_{nm} \langle \sigma_{n}^\dagger\sigma_{m}(\tau)\rangle$ where $\gamma_{nm}$ are the inter-emitter decay rates when $n\neq m$ and single-emitter decay rates associated with the $n$-th emitter otherwise~\cite{ficek2005quantum}. Similar to Sec.~\ref{sec:1e}, the initial dynamics will again be dominated by the linear emitter-phonon coupling, now given by the Hamiltonian $H_{\rm NM}^{\rm 2e} = \sum_{n=1}^2 \left(H_0^{(n)} + H_{\rm e-pn}^{(n)}\right) + H_{\rm pn}$. Then we can express the total correlation as (see Appendix~C),
\begin{equation}
  g^{(1)}(\tau)= 
   \mathcal{C}(\tau)\sum_{n,m=1}^2\gamma_{nm}\rho_{nm}e^{-i\omega_0' \tau},  \;\tau \leq \tau_{\rm P}, 
\label{corr_nonmark_2e}
\end{equation}
where $\rho_{nm} = \langle n | \rho(0) | m\rangle$, with $|n\rangle$ representing the state in the single excitation subspace.\par
After the non-markovian phonon correlations have reached a steady state, we next consider in our concatenation method only the markovian effects due to the photon and photon baths, given by the Hamiltonian $H_{\rm MK}^{\rm 2e} = \sum_{n=1}^2 (H_0^{(n)} + H_{\rm e-pt}^{(n)} + H_{({\rm e-pn})_2}^{(n)} ) + H_{\rm pt} +H_{\rm pn}$. Similar to the single-emitter case, the emitter-photon and the quadratic phonon interactions can be accurately described by a markovian master equation, which for the two identical emitters can be written as ${d\rho}/{dt}= -i[H_s,\rho] +  \gamma_+\mathcal{D}_{\sigma_+}(\rho) + \gamma_-\mathcal{D}_{\sigma_-}(\rho) + 2\gamma_{\rm pd}(T)(\mathcal{D}_{\sigma_+^\dagger\sigma_+}(\rho)+\mathcal{D}_{\sigma_-^\dagger\sigma_-}(\rho))$. This master equation is in diagonal form and describes both collective photon emission and dephasing due to local phonon baths~\cite{ficek2005quantum, Reitz2022CooperativePlatforms}. The local phonon correlations due to an excited emitter become uncorrelated at $\tau > \tau_{\rm P}$ since $\mathcal{C}(\tau) \rightarrow \mathcal{C}_\infty(T)$. This gives a phonon correlation length $L_{\rm P} = c\tau_{\rm P}$, where $c$ is the speed of sound. Since we work in the single-excitation regime and we chose the inter-emitter separations $r_{12} > L_{\rm P}$, the polaron formation occurs on a shorter time than the travel time for phonons to adjacent emitters. Therefore we assume uncorrelated inter-emitter phonon correlations implying independent phonon baths. The diagonalization of the master equation was achieved by using the symmetric operator $\sigma_+ = (\sigma_1 + \sigma_2)/\sqrt{2}$ and its antisymmetric counterpart $\sigma_- = (\sigma_1 - \sigma_2)/\sqrt{2}$. The collective emission is characterized by the superradiant (subradiant) decay rates $\Gamma_+ = \gamma_+/2 + \gamma_{\rm pd}(T)$ ($\Gamma_- = \gamma_-/2 + \gamma_{\rm pd}(T)$), where $\gamma_{\pm} = \gamma \pm \gamma_{\rm col}$ since  we assumed the emitters to be identical and in a bulk environment, they have identical decay rates, i.e., $\gamma_{11} = \gamma_{22} \equiv \gamma$, while $\gamma_{12} = \gamma_{21}\equiv \gamma_{\rm col}$ is the inter-emitter decay rate. The factor $\mathcal{C}(\tau)$ in the correlation function~\eqref{corr_nonmark_2e} incorporates the coherent evolution of emitters influenced by a linear coupling of phonon modes at a picosecond timescale where the markovian effects due to phonons and photons are irrelevant. This initial evolution does not change emitter populations but leads to phonon-induced decoherence. On the other hand, the effect of the photon bath at $\tau > \tau_{\rm P}$ in this two-emitter configuration induces not only decay but also results in frequency renormalization, called the Lamb shift, which is reflected in the coherent part of the markovian master equation via the Hamiltonian $H_s = \omega_{\rm col}( \sigma_+^\dagger\sigma_+ - \sigma_-^\dagger\sigma_-)$, where $\omega_{\rm col}$ is the collective Lamb shift. For bulk GaAs, the inter-emitter decay rates $\gamma_{\rm col}$ and collective Lamb shifts $\omega_{\rm col}$ depend strongly on the distance 
$r_{12}$ between the embedded emitters; in an H-aggregate configuration, where the parallel dipoles are orthogonal to $\mathbf{r_{12}}$, they are given by $\gamma_{\rm col}/\gamma \propto \sin\vartheta/\vartheta + \cos\vartheta/\vartheta^2 - \sin\vartheta/\vartheta^3$ and $\omega_{\rm col}/\gamma \propto -\cos\vartheta/\vartheta + \sin\vartheta/\vartheta^2 + \cos\vartheta/\vartheta^3$. Here, $\vartheta = \mbox{n}q_0' r_{12}$ with $q_0'$ the wavevector associated to the phonon-renormalized emitter resonance frequency $\omega_0'$, and $\rm n$ is the refractive index of GaAs~\cite{Trebbia2022TailoringEmitters, ficek2005quantum}. The eigenstates associated with these two-emitter configurations in the diagonal representation are $\omega_0'\pm \omega_{\rm col}$.\par
We once again concatenate the non-markovian dynamics of~\eqref{corr_nonmark_2e} with the markovian dynamics that is calculated by the QRT, to arrive at the first-order correlation function at all times (see Appendix~C for details)
\begin{equation}
 g^{(1)}(\tau) = \mathcal{C}(\tau)\sum_{l\in (+, -) } \gamma_l\rho_{ll} \mbox{e}^{-(\Gamma_l + i\omega_l' )\tau},\;\;\;\forall\tau,    
  \label{2e_corr_tot}
\end{equation}
where $\omega_{l}' = \omega_0' \pm \omega_{\rm col}$ and $\Gamma_l = \Gamma_\pm$. The calculation of \eqref{2e_corr_tot} is made in the collective basis where $\rho_{ll} = \langle \pm |\rho(0)| \pm\rangle$, with $|+\rangle = (|e,g\rangle + |g,e\rangle)/\sqrt{2}$ is the collective symmetric and $|-\rangle = (|e,g\rangle -  |g,e\rangle)/\sqrt{2}$ the collective antisymmetric state. Since the  collective effects manifest themselves in the $\mu$eV range around the ZPL, while phonon sidebands are spread across the meV range, we can suppress these sidebands with a similar procedure as before for the single emitter, which transforms the concatenation scheme to the simpler initial-slip scheme~\cite{Suarez1992MemorySystems, Cheng2005MarkovianSystems}. This aids in deducing an analytical expression for the emission spectrum near the ZPL in the two-emitter configuration, 
 \begin{equation}
  S_{\rm ZPL}^{\rm 2e}(\omega) =\mathcal{C}_{\infty}(T)\sum_{l \in (+, -)} \gamma_l\rho_{ll}(t)\Gamma_l/\left((\omega - \omega_l')^2 + \Gamma^2_l\right),
  \label{es_2e_zpl}
 \end{equation}
where we sum over the symmetric and antisymmetric eigenstates. The spectrum~\eqref{es_2e_zpl} clearly will show the effects of collective emission, while also reflecting non-markovian behavior of the phonon bath.\par
Now, turning to the polaron method, we subject the two-emitter Hamiltonian $H^{\rm 2e}$ to the unitary transformation $\tilde{H}^{\rm 2e} = \mbox{exp}(S) H^{\rm 2e} \mbox{exp}(-S)$ where $\mbox{exp}(S) = \mbox{exp}(\sum_{n = 1}^2 S_n)$, and $S_n = (g_k^{(n)}b^\dagger - g_k^{*(n)}b)/\omega_k$, to set the stage to derive a polaron master equation (PE) (see Appendix~D for derivation). From $\tilde{H}^{\rm 2e}$ we can calculate the first-order correlation function as $g^{(1)}(\tau) = \sum_{n,m = 1}^2\mathcal{C}_{nm}(\tau)\gamma_{nm}\langle \sigma_n^\dagger\sigma_m(\tau)\rangle_{\rm PE}$, where the two-time correlation function is evaluated with the help of the polaron master equation. Here $\mathcal{C}_{nm}(\tau)$ is the phonon bath correlation function defined as $\mathcal{C}_{nm}(\tau) = \mathcal{C}(\tau)$ when $n = m$ and $\mathcal{C}_\infty(T)$ when $n \neq m$~\cite{Rouse2019OptimalEnvironment}. Expressing $g^{(1)}(\tau)$ in the diagonal form using the symmetric ($\sigma_+$) and antisymmetric operators ($\sigma_-$) as before, while neglecting the sidebands, we obtain $g^{(1)}(\tau) = \mathcal{C}_\infty(T)\sum_{l\in(+, -)}\gamma_{l}\langle \sigma_l^\dagger\sigma_l(\tau)\rangle_{\rm DPE}$, where now the correlation is evaluated using the diagonal form of the two-emitter polaron master equation (DPE) 
\begin{multline}
\frac{d\rho}{dt}= -i[H_{\rm P },\rho] +  \Upsilon_+\mathcal{D}_{\sigma_+}(\rho) + \Upsilon_-\mathcal{D}_{\sigma_-}(\rho)\\
+ 2\gamma_{\rm pd}(T)\left(\mathcal{D}_{\sigma_+^\dagger\sigma_+}(\rho)+\mathcal{D}_{\sigma_-^\dagger\sigma_-}(\rho)\right).    
\label{2e_DPE}
\end{multline}
Here $\Upsilon_\pm = \gamma \pm \mathcal{C}_\infty(T)\gamma_{\rm col}$ and $\Omega_{\rm col} = \mathcal{C}_\infty(T) \omega_{\rm col}$ are the phonon-renormalized temperature-dependent collective decay rate and Lamb shift respectively (see Appendix~E for details) and $H_P =  \sum_{l\in (+, -) } \Omega_l'\sigma_l^\dagger\sigma_l$ where $\Omega_{\pm}' = \omega_0' \pm \Omega_{\rm col}$ are the eigenenergies. Here, one can notice that the single emitter decay rate is not influenced by the phonon bath, which is a consequence of the markovian approximation in the photon bath. On the other hand, the inter-emitter decay rates do exhibit a non-additive phonon influence, see Appendix~E. Thus we can evaluate the first-order correlation function through the QRT utilizing~\eqref{2e_DPE}, which results in the ZPL contribution to the spectrum as
\begin{equation}
 S_{\rm ZPL}^{\rm 2e}(\omega) =\mathcal{C}_{\infty}(T)\sum_{l \in (+, -)} \gamma_l\rho_{ll}(t)\digamma_l\big/\left((\omega - \Omega_l')^2 + \digamma^2_l\right),
\label{2e_es_pm}
\end{equation}
where $\digamma_l = \Upsilon_l/2 + \gamma_{\rm pd}(T)$. This concludes the derivation of the two-emitter spectra in the concatenation and the polaron approaches, where in both cases we provided analytical approximations for the ZPL parts.

 \subsubsection{Comparison of concatenation and polaron methods}
\label{sec:2e_comp}
In this section, we investigate the differences in the two approaches (concatenation scheme and polaron method) in the ZPL spectra by using the combined initial state of the two atoms as $|\psi(0)\rangle = |e,g\rangle$, a localized excitation which equivalently can be written as $|\psi(0)\rangle = 1/\sqrt{2}(|+\rangle + |-\rangle)$. This choice of initial state will result in a contribution from both terms in the emission spectrum in~\eqref{es_2e_zpl} (\eqref{2e_es_pm}), showing up as two distinct peaks, a higher-energy superradiant and a lower-energy subradiant peak, separated by the collective Lamb shift, $\omega_+' - \omega_-' = 2\omega_{\rm col}$ ($\Omega_+' - \Omega_-' = 2\Omega_{\rm col}$). The population dynamics corresponding to this initial state manifests as bi-exponential decay, as illustrated in Appendix~F.
 \begin{figure}
 \centering
    \includegraphics[width = 0.48\textwidth]{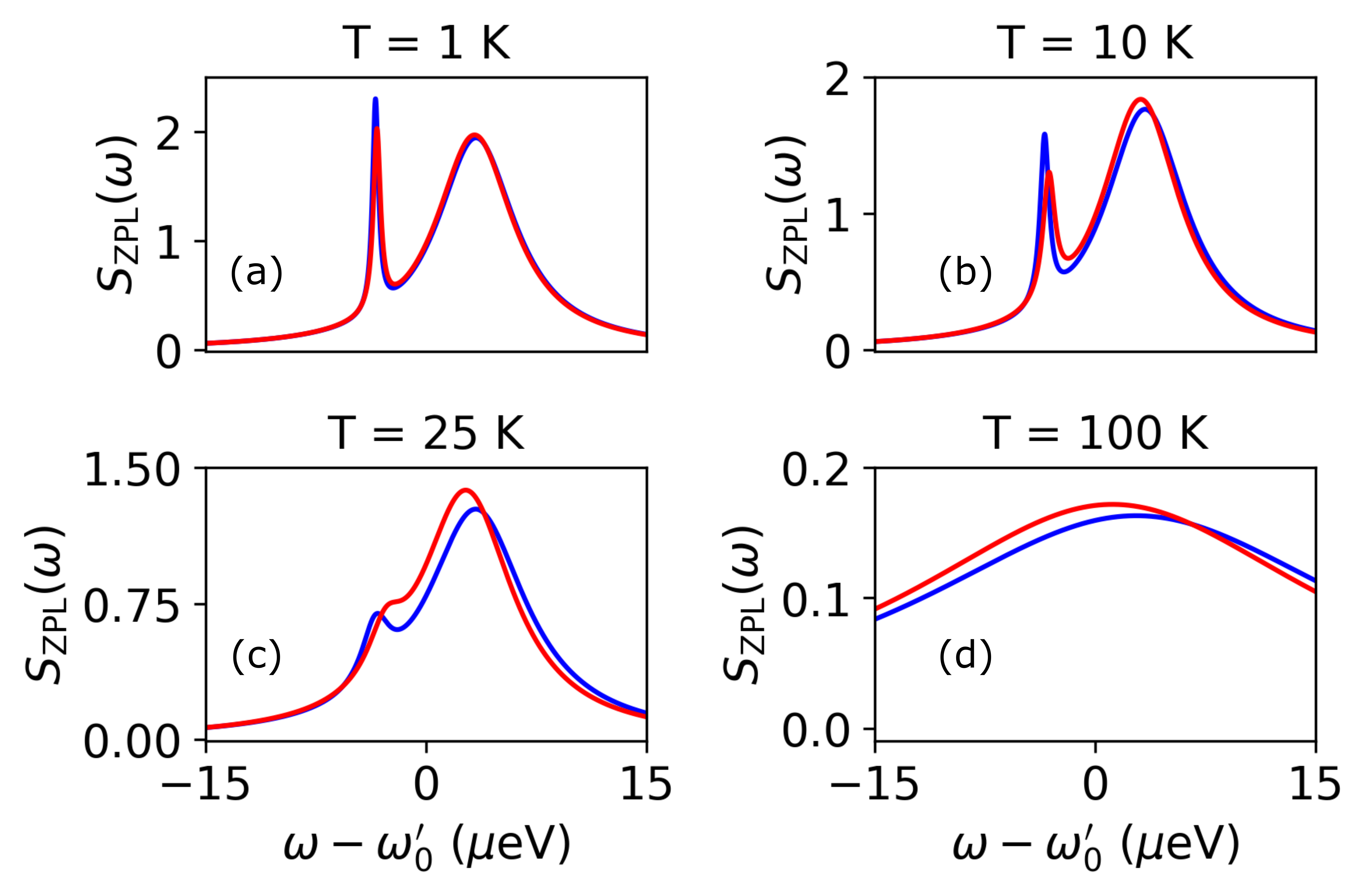}    
    \centering
    \caption{\small{Temperature-dependent emission spectra for two identical emitters prepared in the initial state $|\psi(0)\rangle = |e,g\rangle$, separated by a distance of $\lambda/25$, calculated using the concatenation approach (in blue) and the polaron approach (in red).  Simulation parameters: refractive index $\rm n = 3.5$; $\lambda = 940$ nm; $\gamma = 850$ MHz; $\alpha = 0.025$; $\omega_{\rm c} = 1.49 \rm ps^{-1}$ and the size of the quantum dots is $4.5$ nm. } }
    \label{fig:es_2e}
\end{figure}
 
In Fig.~\ref{fig:es_2e} we depict the ZPL emission spectra based on~\eqref{es_2e_zpl} (\eqref{2e_es_pm}) for four different temperatures and $r_{12} = \lambda/25$ (see Appendix~F for spectrum with $r_{12} = \lambda/15$). We choose these short distances for bulk GaAs with the aim to see a clear manifestation of superradiant and subradiant emission; to observe similar spectra for longer distances, waveguide geometries can be used instead~\cite{Tiranov2023CollectiveEmitters}. The widths of the peaks in Fig.~\ref{fig:es_2e} correspond to superradiant and subradiant decay rates, given by $\Gamma_+$ and $\Gamma_-$ ($\digamma_+$ and $\digamma_-$). A very close agreement between the two approaches can be observed. The agreement is best at the lowest and highest temperatures shown, while at intermediate temperatures, the agreement is still good. The decay rate enhancement $\Gamma_+/\Gamma_-$ can be used to quantify the collective emission~\cite{Tiranov2023CollectiveEmitters}, and it has the value of $13.1 \approx \digamma_+/\digamma_-$, at $T = 1$ K and approaches unity at high temperatures, as shown in Fig.~\ref{fig:es_2e}. The emission spectra obtained with the two methods are different, because the collective decay rate $\Upsilon_{\rm col}(T)$ and collective Lamb shift $\Omega_{\rm col}(T)$ in the polaron approach are temperature dependent due to phonon renormalization, while such a renormalization is not accounted for in the concatenation approach (see Appendix~E and Ref.~\cite{Rouse2019OptimalEnvironment}). This results in slightly different widths of the emission peaks and their respective resonance positions.\par
\begin{figure}
    \centering
    \includegraphics[width = \columnwidth]{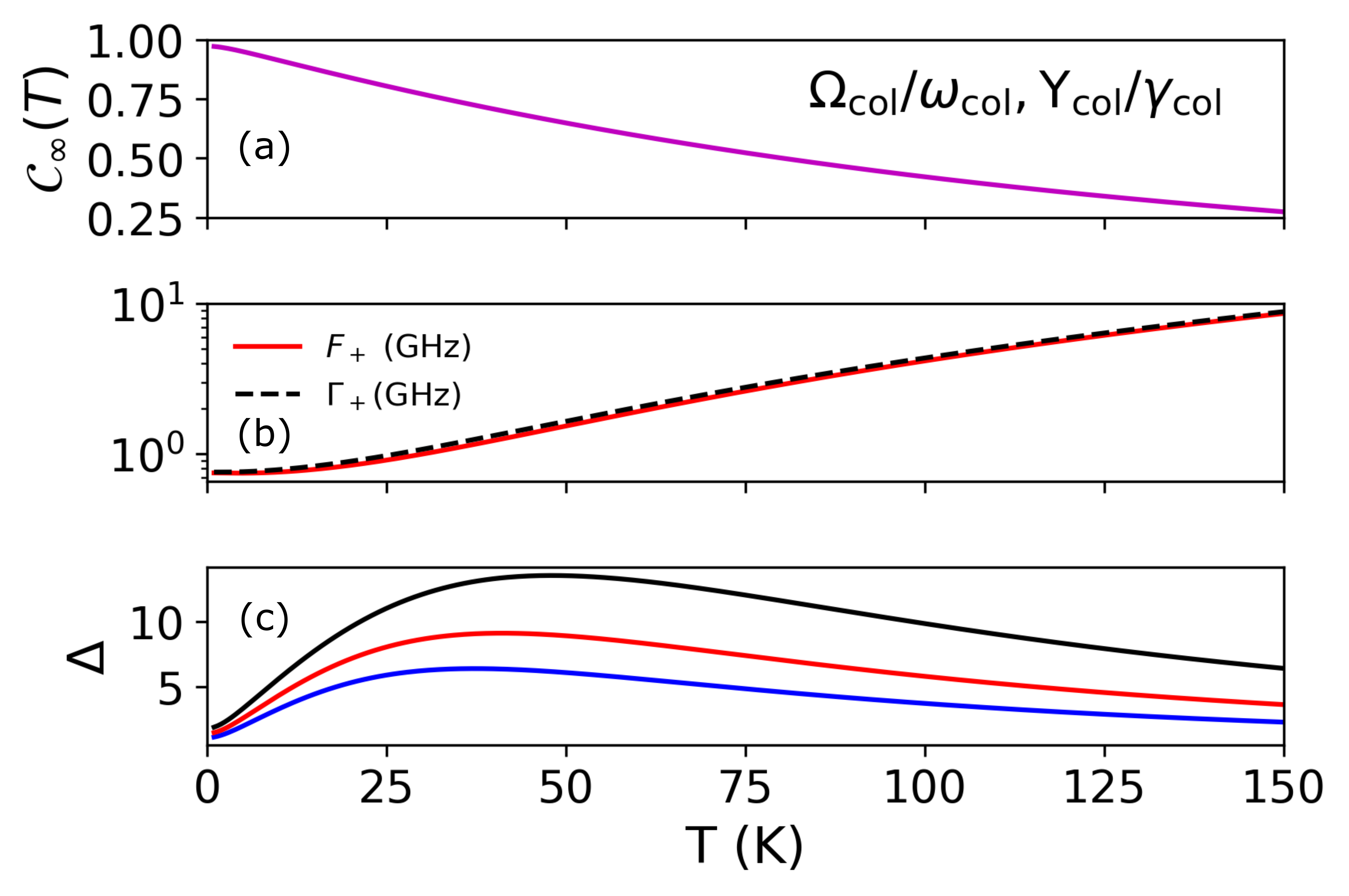}
    \caption{\small{(a) Temperature dependence of the collective Lamb shift (collective decay rate) obtained from the polaron method normalised with the collective Lamb shift (collective decay rate), obtained within the concatenation scheme, for two emitters separated by distance $0.05\lambda$. The ratio is equivalent to the linear coupling factor $\mathcal{C}_{\infty}(T)$. (b) Magnitudes of the superradiant decay rate $\digamma_+$ and $\Gamma_+$ obtained from the polaron theory and our concatenating scheme for different temperatures. (c) The percentage difference between the two schemes $\Delta$ for emitter separations $\lambda/25$ (black line), $\lambda/20$ (red) and $\lambda/15$ (blue).}}
    \label{fig:Tvs}
\end{figure}

We will now quantify how much the ZPL spectra obtained with two methods agree. Fig.~\ref{fig:Tvs}(a) depicts the ratio of the polaron-renormalized collective emission rates ($\Omega_{\rm col}$ and $\Upsilon_{\rm col}$) and the collective rates obtained from the concatenation scheme ($\omega_{\rm col}$ and $\gamma_{\rm col}$), which is equal to $\mathcal{C}_\infty$(T), which therefore measures the variation in the resonance position and the width of the resonance peaks in the two methods. Fig.~\ref{fig:Tvs}(b) shows very similar overall decay rates in both methods: since  $\mathcal{C}_\infty(T)$ differs negligibly from unity at very low temperatures, while at the highest temperatures the markovian phonon pure dephasing dominates, with the same value for $\gamma_{\rm pd}(T)$ in both methods, we find only small differences between the overall decay rates $\Gamma_{\pm}$ and $\digamma_{\pm}$ at all temperatures. We can also define a relative difference between the two methods as $\Delta \equiv 100(\rm CS - PM)/CS$, where $\mbox{CS}  = \Gamma_+ + \Gamma_- + 2\omega_{\rm col}$ constitutes parameters deduced from the concatenation scheme while $\mbox{PM} = \digamma_+ + \digamma_- + 2\Omega_{\rm col}$ constitutes the polaron method. CS and PM both describe the sum of the widths of the super and subradiant emission plus the energy difference between their associated resonance energies. In Fig.~\ref{fig:Tvs}(c), we plot $\Delta$ for three different emitter separations. In all three cases, $\Delta$ starts at a low value at low temperatures, culminating in its highest value at intermediate temperatures, and then decreases monotonously for higher temperatures, in agreement with our previous qualitative observations in Fig.~\ref{fig:es_2e}. Also, a larger separation between emitters results in smaller $\Delta$ (see Appendix~F). In summary, the deviations between the ZPL spectra obtained with the concatenation and polaron methods will for most purposes be negligible at all temperatures.

\subsubsection{Consequence of neglecting non-markovian phonon dynamics}
\label{sec:2e_nmk}

Until now we have compared collective emission in two models that both include markovian as well as non-markovian dynamics due to phonon interactions.  
In our modeling, following Refs.~\cite{Muljarov2004DephasingPhonons, Reigue2017ProbingDots, Tighineanu2018PhononDimensionality}, the pure-dephasing rate originated from the markovian approximation to the second-order phonon interaction, while the first-order phonon interaction was responsible for the non-markovian fast initial dynamics. We note in passing that it would take a one-dimensional phonon bath to obtain from the first-order phonon interaction a nonzero pure dephasing rate in the markovian approximation, while this approximation for a higher-dimensional bath leads to a spurious decoherence-free subspace~\cite{Leggett1987DynamicsSystem, MassimoPalma1996QuantumDissipation, Doll2006LimitationSeparation, Doll2007IncompleteStates, Doll2009FastRates}.\par
As remarked in the Introduction, the influence of phonons on collective emission is typically neglected altogether. Or better, if phonon effects are taken into account, then the state of the art is to only include them through a pure-dephasing rate in a master equation. For example,  in the remarkable recent work by Tiranov et al.~\cite{Tiranov2023CollectiveEmitters}, where two distant quantum dots in a photonic crystal waveguide were shown to emit light collectively,  a markovian master equation was utilized where the emission and emitter interaction were described by the electromagnetic Green function, while a phenomenological dephasing rate accounted for the interaction of the quantum dots with phonons. Our models only reduce to such a markovian master equation by neglecting the first-order interaction with phonons, while retaining the second-order interaction. It is therefore not obvious at the outset that this unusual approximation procedure gives accurate results.  
\begin{figure}
     \centering
     \begin{subfigure}
         \centering
         \includegraphics[width=0.48\textwidth]{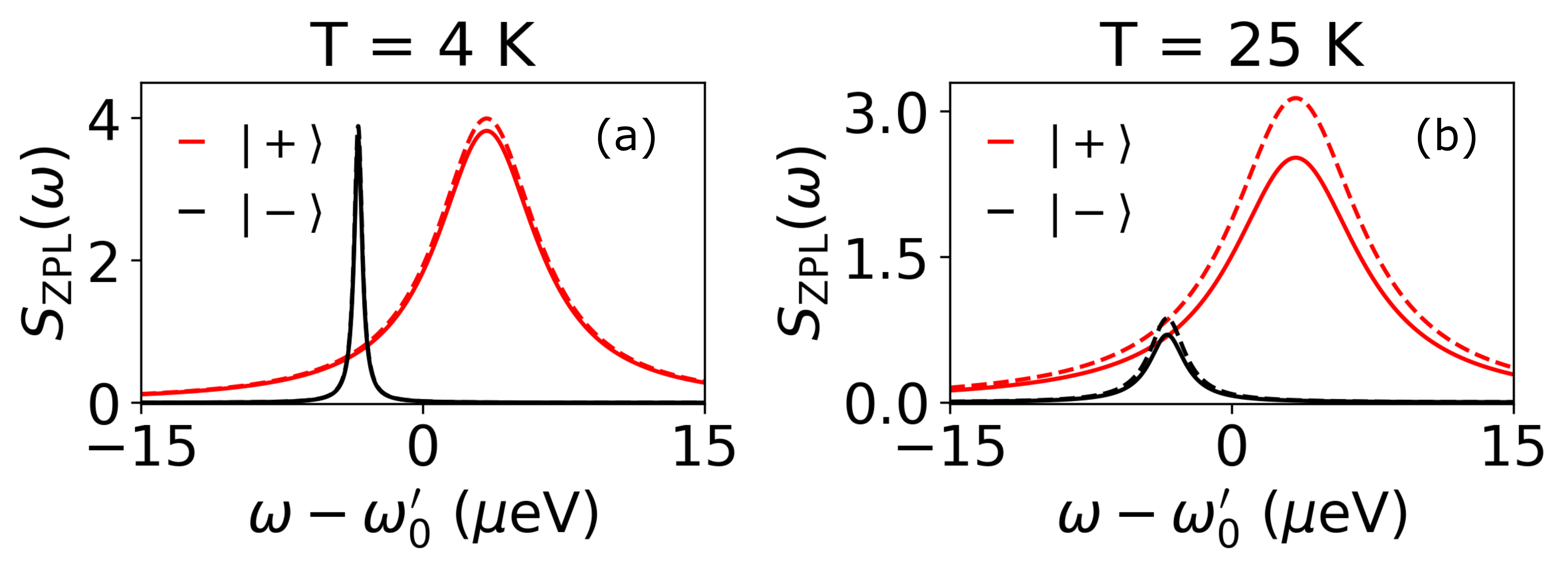}
     \end{subfigure}
     
     \begin{subfigure}
         \centering
         \includegraphics[width=0.48\textwidth]{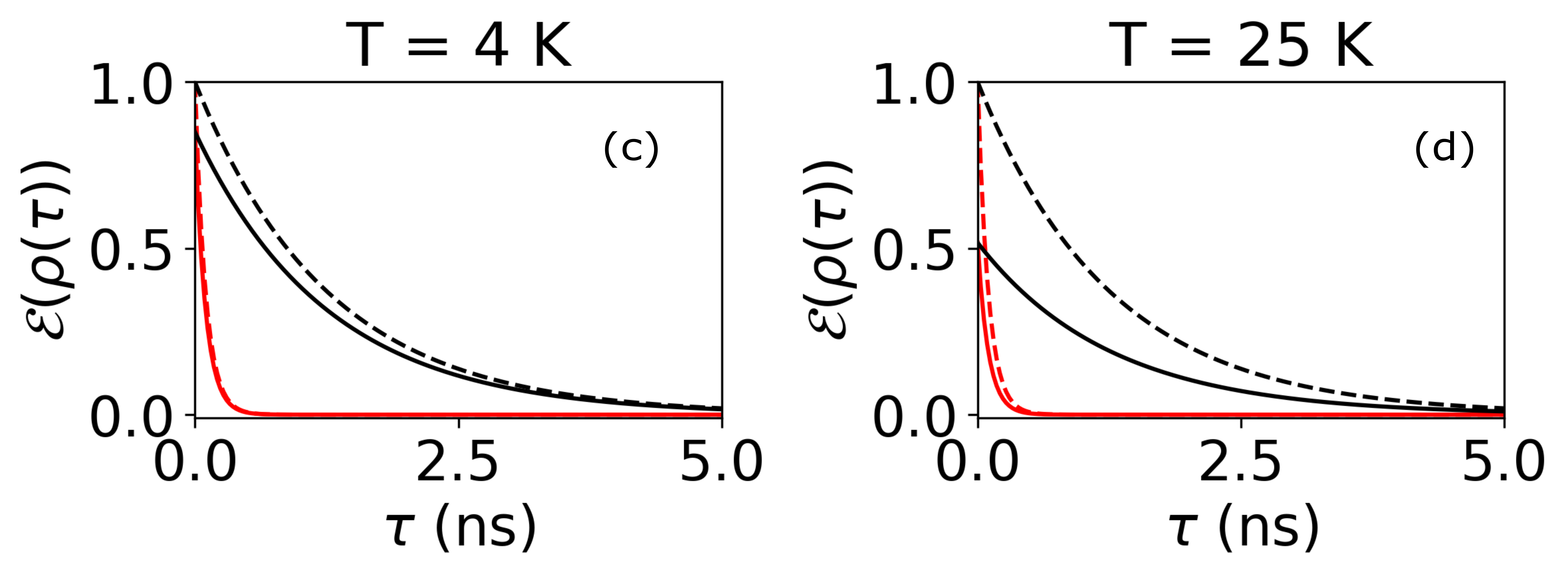}
      \end{subfigure}

      \begin{subfigure}
         \centering
         \includegraphics[width=0.48\textwidth]{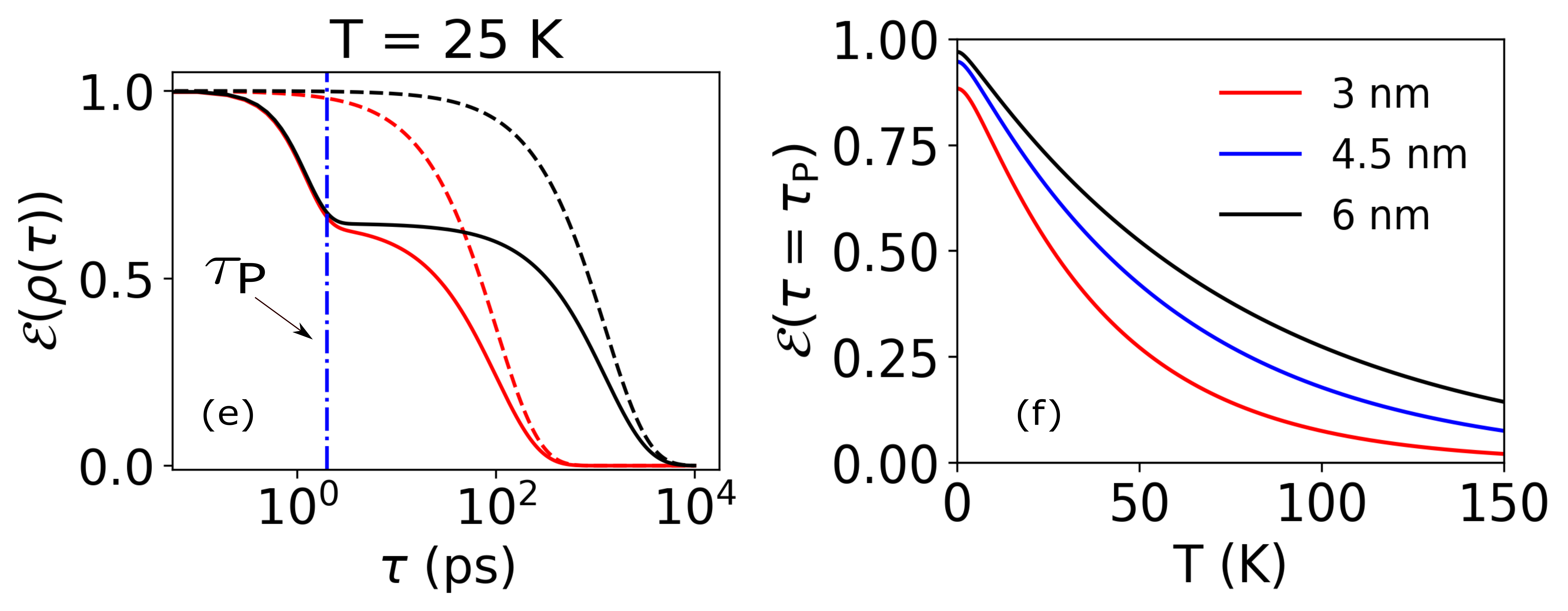}
      \end{subfigure}
      
     \caption{\small{(Top panel) Emission spectrum of Eq.~\eqref{es_2e_zpl} with and without the linear phonon coupling, shown in solid and dashed lines, respectively. Two quantum dots are prepared in the superradiant eigenstate $|+\rangle$ (red curves) and the subradiant state $|-\rangle$ (black), at a temperature of~(a)  4 K and (b) 25 K. The solid and dashed curves coincide in panel~(a). Panels~(c) and (d) depict the corresponding concurrence dynamics. In panel~(e) we show the same data as in panel~(d) but now with the time axis on  a log scale to capture the picosecond dynamics of the concurrence. The blue dashed-dotted vertical line indicates the time  $\tau = \tau_{\rm P}$. Panel~(f) depicts the temperature dependence of the concurrence at time $\tau = \tau_{\rm P}$, i.e. after the initial entanglement decay on the picosecond time scale, for several quantum dot sizes, based on Eq.~\eqref{concur_rho}. The curves in (f) apply to both the super- and the subradiant initial states.}}
        \label{fig: conc_spec}
\end{figure}

Therefore, in Fig.~\ref{fig: conc_spec}, we compare sub- and superradiant spectra of our concatenation model disregarding the sidebands, with the corresponding spectra in the markovian model, as before for the two identical quantum dots in bulk GaAs. In more detail, we compare the approximate spectra of~\eqref{es_2e_zpl} for the concatenation model with its markovian approximation, which amounts to setting $\mathcal{C}_{\infty}(T)$ in \eqref{es_2e_zpl} to unity.  We see that at 4K, the non-markovian dynamics can well be neglected in the $\mu$eV-range around the ZPL, while for 25K, the effect of neglecting the non-markovian dynamics is to overestimate the amplitudes of the spectra, while the shapes of the spectra are the same with or without the non-markovian dynamics in this frequency range. Experimental spectra are often shown in arbitrary units, which thereby become insensitive to non-markovian effects around the frequency range where collective effects are probed.\par
So does this confirm that phonon-induced non-markovian dynamics is negligible? The answer is negative: in the first place, because on a larger (meV) range sidebands, the shapes of the spectra become different with or without non-markovian dynamics as will be shown in Fig.~\ref{fig: conc_spec1} below. In the second place, collective emission may be used as a signature of entanglement~\cite{Tiranov2023CollectiveEmitters, Lohof:2023a}, but with quantum information applications in mind, the actual interest may be in the entanglement dynamics itself.\par
Therefore, as the measure of two-qubit entanglement, we study concurrence, defined as $\mathcal{E}(\rho) = \mbox{max}\{0, \sqrt{\alpha_1} - \sqrt{\alpha_2} - \sqrt{\alpha_3} -\sqrt{\alpha_4}\}$ where $\alpha_i$ are the eigenvalues in descending order of their magnitude, of the matrix $\rho\sigma_{\rm y}^{(1)}\otimes \sigma_{\rm y}^{(2)}\rho^*\sigma_{\rm y}^{(1)}\otimes \sigma_{\rm y}^{(2)}$, with $\sigma_{\rm y}^{(i)}$ the Pauli matrix related to the $i$-th qubit~\cite{Wootters1998EntanglementQubits, Doll2006LimitationSeparation}.
In our concatenation model, the system density matrix at the time $\tau = \tau_{\rm P}$  is given by 
\begin{multline}
\rho(\tau_{\rm P}) = \sum_{n = 0}^2|n\rangle\langle n|\rho_{nn} + \mathcal{C}_\infty^2(T) \sum_{n \neq m = 1}^2 |n\rangle\langle m|\rho_{nm},  
\label{concur_rho}
\end{multline}
where $|0\rangle = |g,g\rangle$ is the collective ground state (see Appendix~G for the derivation), which then serves as the initial state for the markovian master equation for time $\tau > \tau_{\rm P}$. Notice the quadratic dependence of the coherences on $\mathcal{C}_\infty(T)$, which also entails that the concurrence will depend sensitively on any fast initial non-markovian dephasing. In the fully markovian model, again $\mathcal{C}_\infty(T)$ is taken to be unity.\par
In Fig.~\ref{fig: conc_spec}(c,d), we depict the two-emitter entanglement dynamics for the initial states $|+\rangle $ and $|-\rangle$ corresponding to the temperature-dependent spectra in the first two panels. We compare the concatenation model of \eqref{2e_corr_tot} with the fully markovian model. We can see that discarding the phonon-induced non-markovian effects results in overestimating the concurrence. It is a small effect at 4K, which is the operating temperature of Ref.~\cite{Tiranov2023CollectiveEmitters},  but a large non negligible effect already at 25 K. 

The fast initial drop of the concurrence is hard to see in the dynamics on the nanosecond scale in panels 3(c,d). To better visualize this,
in Fig.~3(e) we once more depict the concurrence dynamics at T = 25 K, but now with the time axis on a log scale. We observe sharp non-markovian phonon-induced decay of the concurrence on the picosecond scale, until the phonon bath relaxation time $\tau_{\rm P}$, after which the decay becomes markovian. A full markovian model on the other hand describes only the exponential decay of concurrence, not the initial drop. Lastly, in Fig.~3(f) we depict the temperature-dependent decay in the concurrence at $\tau = \tau_{\rm P}$, for both super- and subradiant initial states, for three sizes of quantum dots. Clearly, the initial decay is larger at higher temperatures and for smaller quantum dots. Smaller quantum dots have larger associated cutoff frequencies $\omega_c$~\cite{Nazir2016ModellingDots, Tighineanu2018PhononDimensionality} and hence  larger initial decay of the concurrence. Keeping the initial decay of concurrence below a few percent requires temperatures of at most a few K. The magnitude of the concurrence at $\tau = \tau_{\rm P}$ is governed by the quadratic dependence on the Franck-Condon factor, see Eq.~(10).

Finally, we conclude this section by comparing the full spectra $S(\omega)$, i.e. also including the sidebands, obtained by numerical simulations in Fig.~\ref{fig: conc_spec1} pertaining to different methods discussed in this section. We find that the concatenation scheme and the polaron method give very similar spectra, where the largest (but still small) differences occur around the ZPL, the frequency interval of Fig.~\ref{fig:es_2e}, where the collective effects manifest themselves. The figure also illustrates that the concatenation and the initial-slip schemes indeed coincide around the ZPL. On the other hand, employing the markovian master equation not only fails to reproduce the sidebands but also may overestimate the amplitude of the spectra around the ZPL as clearly demonstrated in  Fig.~\ref{fig: conc_spec}(b).  
\begin{figure*}
\centering
\includegraphics[width = 0.95\textwidth]{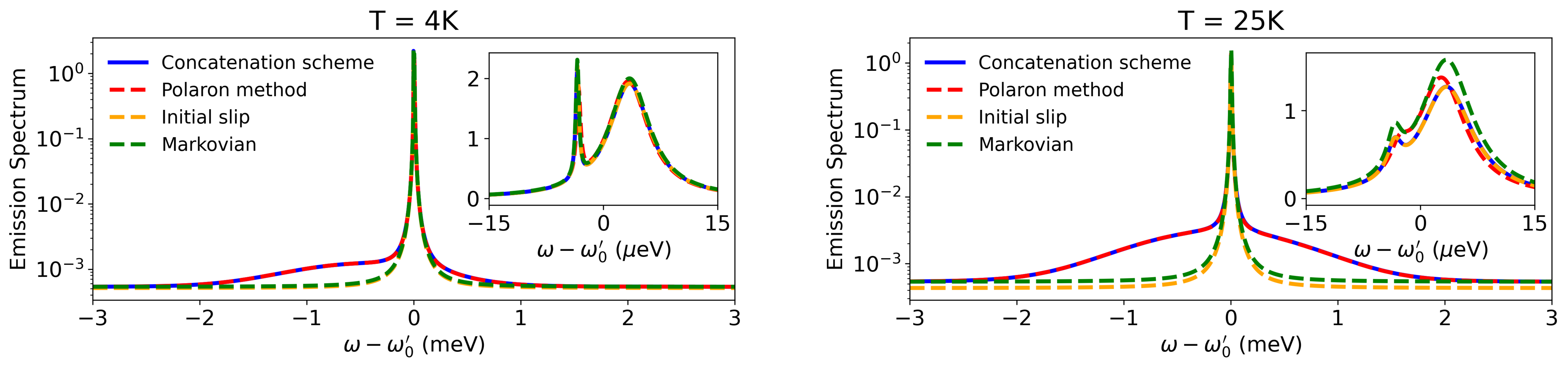}
\caption{\small{(a) Comparison of emission spectra computed with the concatenation model (solid blue line), the polaron method (dashed red line), the initial slip scheme (dashed yellow line) and the markovian master equation (dashed green line) at T = 4 K for two emitters prepared at an initial state $|\psi(0)\rangle = |e,g\rangle$ with emitters separated by distance $\lambda/25$. The range of frequencies in the main panel is in the meV range, depicting the asymmetry in sidebands while the y-axis is on a logarithmic scale. The inset shows a zoomed image of the same plot in the $\mu$eV range, depicting the super- and subradiant splitting with the y-axis on a linear scale, for a good comparison with the plots in Fig.~\ref{fig:es_2e}. (b) A similar plot as in panel (a) but for T = 25 K, resulting in symmetrical sidebands.}}
        \label{fig: conc_spec1}
\end{figure*}

 \subsection{ Scenario III: $N > 2$}
\label{sec:Ne}
A one-dimensional chain of emitters placed over a photonic structure, for instance, a nanofiber or a photonic crystal waveguide, is a promising platform for observing and engineering collective effects~\cite{Vetsch2010OpticalNanofiber, Hood2016Atom-atomWaveguide, Asenjo-Garcia2017ExponentialArrays,Sinha2020Non-MarkovianEmitters}. In this section, we model collective emission in a 1D chain of $N$ equidistant identical emitters in bulk GaAs, generalising the nomenclature and ideas developed in Secs.~\ref{sec:1e} and~\ref{sec:2e}. Thus, we first model the fast phonon dynamics for $N$ emitters interacting with independent phonon baths given by the Hamiltonian, $H_{\rm NM}^{\rm Ne} = \sum_{n=1}^N (H_0^{(n)} + H_{\rm e-pn}^{(n)}) + H_{\rm pn}$ and obtain an exact time evolution, for time $\tau\leq \tau_{\rm P}$. From this exact solution, we obtain the first-order correlation function that captures the non-markovian phonon dynamics as follows (see Appendix~C for details),
\begin{equation}
  g^{(1)}(\tau) = \mathcal{C}(\tau)\sum_{n,m=1}^N\gamma_{nm}\rho_{nm}e^{-i\omega_0' \tau},  \;\tau \leq \tau_{\rm P},
  \label{nmk_NE}
\end{equation}
where $\rho_{nm} = \langle n |\rho(0)| m \rangle$, with the states, $|n\rangle \equiv |g , g,..., e_n,..., g\rangle$ being the localised states, where only the $n$-th emitter is excited, together spanning the single-excitation manifold, while $\gamma_{nm}$ are the inter-emitter decay rates. For $\tau > \tau_{\rm P}$ we can once again use the Markovian description described by the Hamiltonian $H_{\rm MK}^{\rm Ne} = \sum_{n=1}^N (H_0^{(n)} + H_{\rm e-pt}^{(n)} + H_{({\rm e-pn})_2}^{(n)} ) + H_{\rm pt} +H_{\rm pn}$, analogous to Sec.~\ref{sec:2e}. From this Hamiltonian, one obtains the first-order correlation function of the form $g^{(1)}(\tau) = \sum_{n,m=1}^N\gamma_{nm} \langle \sigma_{n}^\dagger\sigma_{m}(\tau)\rangle_{\rm ME}$, where the inter-emitter decay rates $\gamma_{nm}$ satisfy $\gamma_{n,n+1} = \gamma_{\rm col}$ and $\gamma_{n,m} = \gamma_{m,n}$. The correlation $\langle\cdot\cdot\cdot\rangle_{\rm ME}$ will be evaluated with the aid of the QRT utilizing the multi-emitter markovian master equation (ME), which we will derive in the nearest-neighbour interaction approximation in order to arrive at some analytical results. (It is not necessary to make the approximation for our numerical schemes to work, though.) As the name already indicates, in the nearest-neighbour interaction approximation only the dipole-dipole interaction $\omega_{nm}$ with nearest neighbours are considered, i.e. $\omega_{nm} = \omega_{\rm col}$ if $|n - m| = 1$ and zero otherwise, where $n$ and $m$ label the $n$-th and $m$-th emitters, respectively~\cite{Reitz2022CooperativePlatforms}. Such a master equation written in terms of atomic operators $\sigma_n$ is given by
\begin{multline}
   \frac{d\rho}{dt} = -i\sum_{n=1}^N\bigg[\omega_{\rm col}(\sigma_n^\dagger\sigma_{n + 1} + \sigma_{n + 1}^\dagger\sigma_{n}), \rho\bigg]\\ +  \sum_{n,m=1}^N\gamma_{nm}\mathcal{D}_{\sigma_{nm}}(\rho) + 2\gamma_{\rm pd}(T)\sum_{n=1}^N\mathcal{D}_{\sigma_n^\dagger\sigma_n}(\rho),
   \label{me_Ne_main}
\end{multline}
where the dissipator $\mathcal{D}_{\sigma_{nm}}(\rho) = \sigma_n\rho\sigma_{m}^\dagger-\frac{1}{2}\{\sigma_n^\dagger\sigma_{m},\rho\}_+$. 
For analytical as well as numerical calculations, it is convenient to use the diagonal form of such a master equation. For $N = 2$, obtaining a simultaneous exact diagonalisation of the coherent and dissipative parts of the master equation described in Sec.~\ref{sec:2e} was rather intuitive by using the symmetric and antisymmetric combinations of the atomic operators to obtain the collective operators. For $N > 2$, generally, the simultaneous exact diagonalisation of the the coherent and dissipative parts of the master equation is not possible, as outlined in Appendix~C (also see~\cite{Reitz2022CooperativePlatforms}). We will resort to an approximate treatment for both methods to obtain  analytical results.\par
Assuming that each emitter interacts with its own local bulk phonon environment, we can describe the problem in a collective basis in an $N$-dimensional single-excitation manifold.
The states $|n\rangle$ describing localized single-photon excitations can equivalently be written in terms of linear combinations of the collective states, $|l\rangle$ i.e. $|n\rangle = \sum_{l= 1}^N M_{nl}|l\rangle$, where $M_{nl} = \sqrt{2/(N + 1)}\sin\left(\pi n l/ (N + 1)\right)$ are orthonormal coefficients~\cite{Plankensteiner2015, Reitz2022CooperativePlatforms}.  The collective states  $|l\rangle = \sigma_l^\dagger|0\rangle$ where $|0\rangle \equiv |g, g,...,g\rangle$ as the collective ground state and $\sigma_l^\dagger$ is the collective raising operator which excites the collective ground state to the collective state $|l\rangle$. The atomic operators similarly undergo the same transformation, i.e. $\sigma_n = \sum_{l= 1}^N M_{nl}|\sigma_l\rangle$.  Thus, the $N$-emitter master equation can be expressed in terms of the collective operators $\sigma_l$ (details in Appendix~C), leading to a diagonal form of the master equation corresponding to $N$ independent photon decay channels, each one associated with one of the collective states. The resulting eigenstates are given by $\omega_l = 2\omega_{\rm col}\cos\left(\pi l/(N +1)\right)$, with the corresponding collective decay rates $\gamma_l = \sum_{n,m = 1}^N\gamma_{nm} M_{nl}M_{ml}$. In the Dicke limit, i.e. with zero separation between the emitters, $\gamma_{nm} = \gamma$ implying that $\gamma_1\approx \gamma N$ (superradiant decay) and $\gamma_N\approx 0$ (subradiant decay)~\cite{Plankensteiner2015}.\par
The total first-order correlation function in terms of the collective states can be expressed as $g^{(1)}(\tau) = \mathcal{C}_\infty(T) \sum_{l= 1}^N\gamma_{l} \langle \sigma_{l}^\dagger\sigma_{l}(\tau)\rangle_{\rm DME}$, where the correlation $\langle \cdot\cdot\cdot \rangle_{\rm DME}$ is now evaluated using the QRT from the diagonalised multi-emitter master equation 
\begin{equation}
    \frac{d\rho}{dt} = \sum_{l=1}^N\left(-i\left[\omega_l\sigma_l^\dagger\sigma_l,\rho\right] +  \gamma_l\mathcal{D}_{\sigma_l}(\rho) + 2\gamma_{\rm pd}(T)\mathcal{D}_{\sigma_l^\dagger\sigma_l}(\rho)\right).\\
    \label{dme_Ne_main}
\end{equation}
 Concatenating the diagonal form of \eqref{nmk_NE} and the markovian dynamics evaluated via \eqref{dme_Ne_main}, we obtain the first-order correlation function at all times in the collective basis 
\begin{equation}
 g^{(1)}(\tau) = \mathcal{C}(\tau)\sum_{l =1}^N \gamma_l\rho_{ll} \mbox{e}^{-(\Gamma_l + i\omega_l' )\tau},\;\;\;\forall\tau,    
  \label{Ne_corr_CT}
\end{equation}
where $\rho_{ll} = \langle l |\rho(0)| l\rangle$ and $\omega_l' = \omega_0' + 2\omega_{\rm col}\cos\left(\pi l/(N +1)\right)$. The emission spectrum for $N$ emitters neglecting the sidebands evaluated from \eqref{Ne_corr_CT} can therefore be given as
 \begin{equation}
 S_{\rm ZPL}^{\rm Ne}(\omega) =\mathcal{C}_{\infty}(T)\sum_{l = 1}^N \gamma_l\rho_{ll}\Gamma_l/\left((\omega - \omega_l')^2 + \Gamma^2_l\right),
 \label{Ne_CT}
 \end{equation}
 where the total collective decay rate is $\Gamma_l = \gamma_l/2 + \gamma_{\rm pd}(T)$.\par
 In the polaron method we obtain an $N$-emitter polaron master equation (PE) by subjecting the Hamiltonian describing an $N$-emitter phonon and photon interaction via $H^{\rm Ne} = \sum_{n=1}^N \big(H_0^{(n)} + H_{\rm e-pt}^{(n)} + H_{({\rm e-pn})_1}^{(n)} + H_{({\rm e-pn})_2}^{(n)} \big) + H_{\rm pt} + H_{\rm pn}$, to the unitary polaron transformation $\tilde{H}^{\rm Ne} = \mbox{exp}({S}) H^{\rm Ne} \mbox{exp}(-S)$ where now $\mbox{exp}(S) = \mbox{exp}(\sum_{n = 1}^N S_n)$ with $S_n$ defined in Sec.~\ref{sec:2e}. We can thereby obtain the first-order correlation function by utilizing the definition of the phonon bath correlations in Sec.~\ref{sec:2e} and suppressing the sidebands as $g^{(1)}(\tau) = \mathcal{C}_\infty(T)\sum_{n,m = 1}^N\gamma_{nm}\langle \sigma_n^\dagger\sigma_m(\tau)\rangle_{\rm PE}$, where the $N$-emitter polaron master equation (PE) is given by (see Appendix~D for details),
 \begin{multline}
 \frac{d\rho}{dt}= -i\sum_{n=1}^N\bigg[\omega_0'  \sigma_n^\dagger\sigma_n + \Omega_{\rm col}\left(\sigma_n^\dagger\sigma_{n + 1} + \sigma_{n + 1}^\dagger\sigma_{n}\right), \rho\bigg] \\ +  \sum_{n,m= 1}^N \Upsilon_{nm}\mathcal{D}_{\sigma_{n,m}}(\rho) + 2\gamma_{\rm pd}(T)\sum_{n=1}^N \mathcal{D}_{\sigma_{n}^\dagger\sigma_n}(\rho),   
 \label{SPME_nd}
\end{multline}
and where the inter-emitter decay rate $\Upsilon_{nm} = \gamma$ when $n = m$ while $\Upsilon_{nm} = \gamma_{nm}\mathcal{C}_\infty(T)$ when $n \neq m$ (see Appendix~E for details). 
In the diagonal representation the first-order correlation takes the form $g^{(1)}(\tau) = \mathcal{C}_\infty(T) \sum_{l= 1}^N\gamma_{l} \langle \sigma_{l}^\dagger\sigma_{l}(\tau)\rangle_{\rm DPE}$, where the correlations are computed using the multi-emitter diagonal form of the polaron master equation (DPE) 
 \begin{equation}
    \frac{d\rho}{dt} = \sum_{l=1}^N\left(-i\left[\Omega_l'\sigma_l^\dagger\sigma_l,\rho\right] +  \Upsilon_l\mathcal{D}_{\sigma_l}(\rho)
    + 2\gamma_{\rm pd}(T)\mathcal{D}_{\sigma_l^\dagger\sigma_l}(\rho)\right),\\
    \label{DPE_Ne}
\end{equation}
which was obtained from the same diagonalization scheme as was used to diagonalize~\eqref{me_Ne_main}, and where $\Upsilon_l = \sum_{n,m= 1}^N\Upsilon_{nm} M_{nl}M_{ml}$, and $\Omega_l' = \omega_0' + 2\Omega_{\rm col}\cos\left(\pi l/(N +1)\right)$.  Thus, the first-order correlation function takes the form
\begin{equation}
g^{(1)}(\tau) = \mathcal{C}_\infty(T) \sum_{l =1}^N \gamma_l\rho_{ll} \mbox{e}^{-(\digamma_l + i\Omega_l' )\tau},\;\;\;\forall\tau,    
\label{Ne_corr_tot_main}
\end{equation}
where $\digamma_l = \Upsilon_l/2 + \gamma_{\rm pd}(T)$. From \eqref{Ne_corr_tot_main}, we obtain the emission spectrum
\begin{equation}
 S_{\rm ZPL}^{\rm Ne}(\omega) =\mathcal{C}_{\infty}(T)\sum_{l = 1}^N \gamma_l\rho_{ll}\digamma_l/\left((\omega - \Omega_l')^2 + \digamma^2_l\right).
\label{Ne_es_pm}
\end{equation}
 \begin{figure}
 \centering
    \includegraphics[width = 0.48\textwidth]{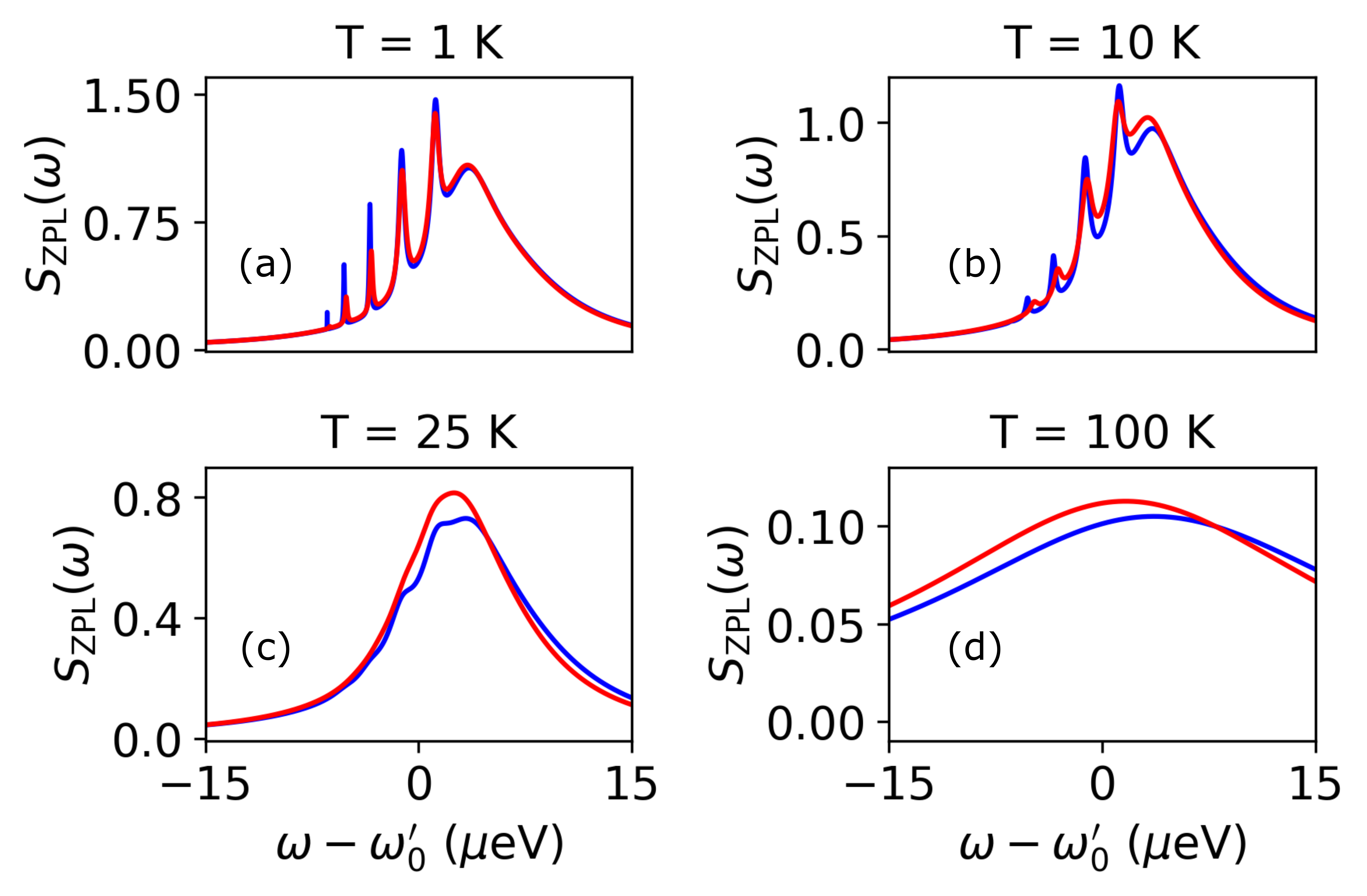}    
    \caption{\small{Emission spectrum for eight equidistant emitters on a line with each emitter separated by a distance of $\lambda/25$ from the other, prepared in the initial state $|\psi(0)\rangle = |e_1 , g, g, g,...\rangle$, manifesting collective emission at different temperatures using our concatenation approach (in blue) and the polaron approach (in red) for different temperatures. }}
    \label{fig:es_8e}
\end{figure}
Choosing $N = 8$ and using as our initial state the localized single-photon state $|\psi(0)\rangle = |e_1 , g, g, g,...\rangle$ results in the emission spectrum depicted in Fig.~\ref{fig:es_8e}, from the concatenation scheme (polaron method) utilizing~\eqref{Ne_CT}  (\eqref{Ne_es_pm}). The two approaches show appreciable agreement for the lowest and highest temperatures, with a maximum deviation at intermediate temperatures that is still small, just as we found for two emitters in Sec.~\ref{sec:2e}.\par
Although for $N=8$ there are eight eigenfrequencies $\omega_l'$ (and $\Omega_l'$), in Fig.~\ref{fig:es_8e}(a) only six distinct resonance peaks can be distinguished since the peaks corresponding to the three highest eigenenergies have merged together. These multi-peak spectral characteristics can be regarded as signatures of collective phenomena. A complete absence of collective effects would result in only one emission peak corresponding to single-emitter decay. A decay-rate enhancement $\Gamma_{\rm max}/\Gamma_{\rm min} = 946 \approx \digamma_{\rm max}/\digamma_{\rm min} $ at T = 1K is deduced, which reduces to 6.85 already at 25K and approaches unity as the temperature increases beyond 100K. This decay rate enhancement quantifies the collective emission and is an indication of entanglement between the multiple emitters. 
 
\section{Conclusions and outlook}
\label{sec:conclusions}

We presented two methods to describe collective light emission in photonic media, which both account for markovian as well as non-markovian dephasing due to interactions with phonons. 
We proposed and compared the concatenation approach and the polaron method, and deduced simple analytical results for an arbitrary number of emitters interacting with a common photon bath and independent phonon baths. The methods give very similar results. While both methods are computationally efficient,  the concatenation approach is the simplest of the two that takes into account the phonon effects without much additional complexity compared to all-optical master equations.\par
Our proposed concatenation scheme concatenates the exact solution of the initial non-Markovian dynamics due to phonons with the subsequent markovian dynamics due to both photon and phonon baths as decribed by a multi-emitter markovian master equation.  Our analysis was carried out for emitters embedded in a bulk GaAs environment. Utilising the bulk phonon spectral density for GaAs allows us to identify an ultrafast cutoff time scale $\tau_{\rm P}$ on which the phonon correlations reach a steady state, the key feature that made our concatenation scheme possible. In the second, markovian part of the dynamics, the effects of the photon and phonon baths are additive, in the sense that they are described by independent exponential decay rates.\par
The other method, the polaron method,  is a state-of-the-art technique that has been utilised to study a single emitter interacting with phonons and photons. We extended it for a multi-emitter 1D chain configuration exhibiting collective effects by deriving a polaron master equation for an arbitrary number of emitters. In general, in the polaron method the optical decay rates are modified by the phonon bath, as an example of the so-called non-additive effects of the two baths. 
However, here we considered the common situations where local optical densities of states vary little across the emitter linewidths (flat spectral density approximation), in which case the non-additivity of single-emitter decay rates disappears, resulting in single-emitter decay rates that are identical to the ones in the concatenation scheme. On the other hand, inter-emitter decay rates (and hence also collective emission rates)  do exhibit the non-additive effects of the phonon bath in the polaron method, whether one makes the flat photon spectral density approximation or not, as a consequence of the assumption that the phonon baths of different emitters are independent. \par
We deduced approximate analytical expressions for the emission spectrum around the zero-phonon line, after employing a known approximate simultaneous diagonalization scheme for the coherent and dissipative parts of the master equation in both methods. We also made a rigorous quantitative comparison of the two methods for different temperatures and found that they agree very well. Thus using the polaron method instead of the simpler concatenation method seems like overkill, at least for the flat photon spectral densities considered here.\par
Decoherence due to the environment of solid-state emitters is often modeled as a single markovian dephasing rate, which from the perspective of our Hamiltonian model amounts to neglecting the linear phonon interaction (non-markovian dephasing) while keeping the quadratic interaction (markovian dephasing). We studied the accuracy of this procedure for various temperatures. As a main result, we found that with a purely markovian model for phonon effects one may considerably overestimate the amount of quantum entanglement (measured as concurrence) between quantum emitters. However, for quantum dots in bulk GaAs at temperatures down to a few degrees Kelvin, the markovian model gave almost the same entanglement dynamics as our concatenation model. Indeed, in the limit of zero temperature, both models reduce to the same well-known all-optical multi-qubit master equations.\par
For simplicity, we considered bulk GaAs as the photonic environment. For appreciable collective emission, we put the emitters at strongly subwavelength distances, much shorter than in the recent experiments in photonic crystal waveguides~\cite{Tiranov2023CollectiveEmitters}. This illustrates that waveguide structures are better suited for observing collective emission at longer distances. Our formalism can be generalized to such inhomogeneous lossless media, by replacing the transverse plane waves with the optical eigenmodes of those media. In most inhomogeneous media that do not involve emitters resonantly coupled to optical cavities, the flat photon spectral density approximation will still hold.\par
In the regime where we work it is natural to assume independent phonon baths to model the collective effects in the presence of phonons. Scenarios where correlated phonon baths as in Refs.~\cite{MassimoPalma1996QuantumDissipation, Reina2002DecoherenceRegisters, Doll2007IncompleteStates, Valido2013GaussianEnvironment, Addis2013Two-qubitEnvironment, Jeske2013DerivationDecoherence} show their marks would introduce another source of non-markovian phonon behaviour, and in the future it would be interesting to study their effects on collective light emission as well. Finally, since there are numerically exact approaches capturing the non-markovian phonon effects and the excitation pulse~\cite{Vagov2002Electron-phononPulses, Barth2016Path-integralSystems, Richter2022EnhancedNetworks, Cygorek2022SimulationEnvironments}, it will be  interesting to  model collective emission and entanglement dynamics with them and to compare them with our methods in several parameter regimes.

\section*{ACKNOWLEDGMENTS}
M.W. and D.P. acknowledge the Independent Research Fund Denmark Natural Sciences (Project No. 0135-00403B). M.W. acknowledges the support by the Danish National Research Foundation through NanoPhoton Center for Nanophotonics, Grant No. DNRF147, and Center for Nanostructured Graphene, Grant No. DNRF103. We thank Mads Anders Jørgensen, Jakob Hummelgaard, Tobias Egebjerg, Nicolas Stenger, Sanshui Xiao and Jake Iles-Smith for stimulating discussions.
\onecolumngrid
\appendix
\renewcommand\thefigure{\thesection.\arabic{figure}}
\section{Correlations and spectrum of an emitter interacting with photon and phonon baths}
In this section, we plot the first-order correlation function in Fig.~\ref{fig1:1ecorr}(a) which illustrates very well our motivation to model the non-Markovian and Markovian dynamics separately. At $\tau < \tau_{\rm P}$, the emitter dynamics is dominated by the non-Markovian dephasing due to the phonon bath, resulting in a decay of the correlations on a picosecond scale. We model this fast decay by the exact solution of the independent boson model. After $\tau_{\rm P}$, the dynamics is dominated by the Markovian decay due to both photon emission and phonon-assisted pure dephasing. The markovian pure dephasing due to phonons, quantified by the dephasing rate $\gamma_{\rm pd}(T)$, is very sensitive to the temperature, as shown in  Fig.~\ref{fig1:1ecorr}(b). This dephasing rate has been calculated using $\gamma_{\rm pd}(T) \equiv\frac{\alpha^2\mu}{\omega_{\rm c}^4}\int_0^\infty \omega^{10}e^{-2\omega^2/\omega_{\rm c}^2}n(\omega)(n(\omega)+1)d\omega$~\cite{Reigue2017ProbingDots, Tighineanu2018PhononDimensionality}, where $\alpha$ is the deformation potential coupling constant, $n(\omega) = 1/(\mbox{exp}(\beta\omega)-1)$ is the occupation number for acoustic phonons with  $\beta = 1/k_{\rm B }T$ the inverse temperature and $k_{\rm B}$ the Boltzmann constant and $\mu = \pi d^4(D_{\rm e}^2m_{\rm e} + D_{\rm h}^2m_{\rm h})^2(D_{\rm e} - D_{\rm h})^{-4} $, where the $m_{\rm e/h}$ are the electron and hole effective masses and, $D_{\rm e}$ and $D_{\rm h}$ are the electron and hole deformation potentials respectively. The temperature dependence of the dephasing rate is a consequence of the change in the phonon occupation $n(\omega)$ with temperature.   
\setcounter{figure}{0}
\begin{figure}[h]
\centering
    \includegraphics[width = 0.7\textwidth]{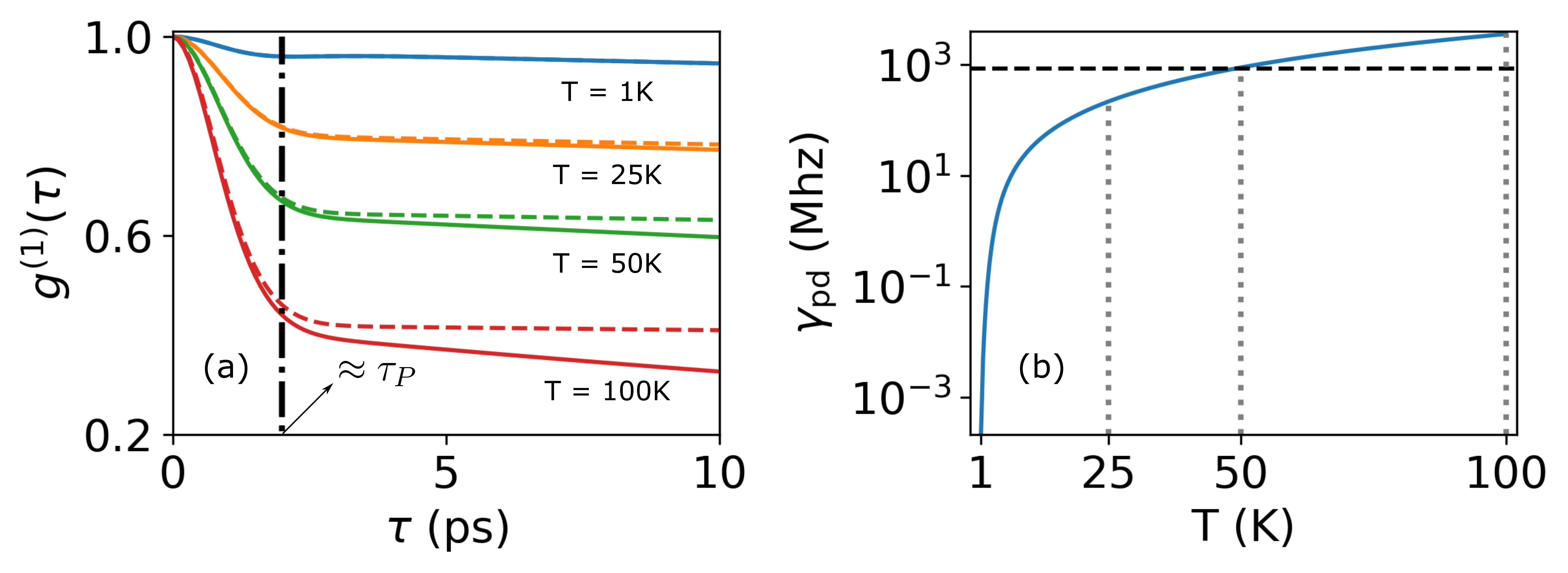}
    \caption{\small{(a) Correlation dynamics for different temperatures taking into account only linear phonon coupling (dashed lines) and both linear and quadratic phonon coupling (solid lines). There is a clear separation between fast and slow dynamics. (b)  The markovian pure dephasing rate, $\gamma_{\rm pd}(T)$ as a function of temperature. The horizontal dashed line shows the magnitude of $\gamma$, and vertical dotted lines point to the corresponding values of $\gamma_{\rm pd}(T)$ for the specific choice of temperatures used to plot the figure on the right panel. Simulation parameters used are $\lambda = 940$ nm; $\gamma = 850$ MHz; $\alpha = 0.025$; $\omega_{\rm c} = 1.49 \rm ps^{-1}$ and the size of the quantum dot $4.5$ nm. All the simulations in the main text and in the supplement material are carried out on the Python-based tool QuTiP~\cite{Johansson2012QuTiP:Systems, Johansson2013QuTiPSystems}. } }
    \label{fig1:1ecorr}
\end{figure}
In Fig.~\ref{fig:es_1e} we also show single-emitter emission spectra obtained with the concatenation scheme, for various temperatures. An important point, stressed in the main text, is that these spectra are identical to the ones obtained with the polaron approach. These single-emitter spectra will be useful for comparison with collective emission spectra for several emitters. The figure depicts the well-known asymmetry in the sidebands at low temperatures, because phonons can be created but are not likely to be absorbed during the photon emission.
At higher temperatures,  both phonon creation or absorption are possible, leading to prominent and symmetric sidebands in the emission spectrum. The inset of Fig.~\ref{fig:es_1e} shows the ZPL broadening due to the thermal phonons and its attenuation by a factor $\mathcal{C}_{\infty}(T)$.
\begin{figure}[h]
\centering
   \includegraphics[width = 0.55\textwidth]{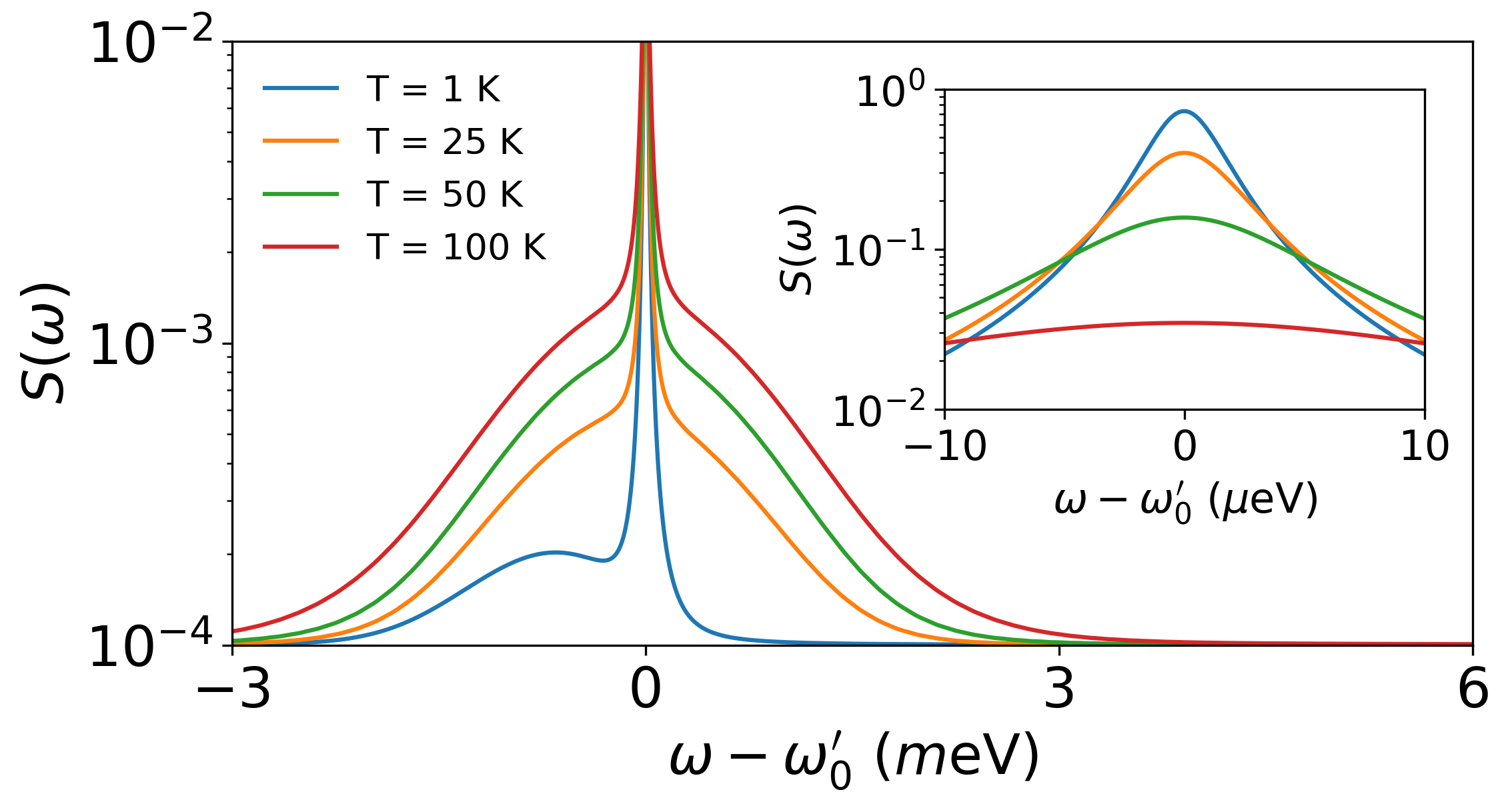}
    \caption{\small{Single-emitter emission spectra for various temperatures, calculated with the concatenation method. Pronounced phonon sidebands are visible at higher temperatures. The inset depicts a magnified image around the zero-phonon line, capturing the broadening of the zero-phonon line due to the combination of spontaneous decay and phonon-induced pure dephasing.  Identical single-emitter spectra are obtained when using the polaron approach instead~\cite{Chassagneux2018EffectRegime, Iles-Smith2017LimitsEmitters, Reigue2017ProbingDots, Nazir2016ModellingDots, Iles-Smith2017PhononSources}.}}
   \label{fig:es_1e}
\end{figure}

\section{Derivation of $g^{(1)}(\tau)$ from concatenating scheme for $N$ = 1}
\label{secii}
The field correlation defined as the first-order correlation function, $g^{(1)}(\tau)$ describes the time dependence at which the coherences of the emitted light evolve. It also leads to the calculation of the emission spectrum. To calculate the correlation we use $\langle E^-(0)E^+(\tau)\rangle = g^{(1)}(\tau) = \gamma\langle \sigma^\dagger\sigma(\tau)\rangle_{\rm NME}$, where the atomic correlations are evaluated by the non-markovian dynamical equation (NME) obtained from the Hamiltonian \eqref{IBM} which is referred to as the independent boson model (IBM) an instance of the exactly solvable independent boson model (IBM)~\cite{Duke1965Phonon-broadenedStates,Mahan2000Many-ParticlePhysics,Krummheuer2002TheoryDots}.
We can also express the correlations as $ g^{(1)}(\tau) = \gamma\mbox{Tr}\left[\sigma^\dagger\sigma(\tau)\chi(0)\right]$, where $\chi(0)$ is the combined system and environment state and we would need to evaluate the temporal evolution of the operator $\sigma(\tau) = U^\dagger(\tau)\sigma U(\tau)$. Here, $U(\tau)$ is the unitary evolution of the system describing the non-markovian fast phonon dynamics, given by the Hamiltonian
\begin{equation}
    H_{\rm NM}^{\rm 1e} = \omega_0\sigma^\dagger\sigma + \sigma^\dagger\sigma\sum_k\big(g_k b_{k}^\dagger +g_k^{*}  b_{k}\big) + \sum_{k}\omega_{k} b_{k}^\dagger b_{k}.
    \label{IBM}
\end{equation}
In~\eqref{IBM} the first term describes the system Hamiltonian for a two-level emitter with the resonance frequency $\omega_0$. Here $\sigma$ ($\sigma^\dagger$) is the lowering (raising) atomic operator and the $b_k$ ($b_k^\dagger$) are phonon annihilation (creation) operators, while $g_k$ is the emitter-phonon coupling constant. From \eqref{IBM} we can see that $[\sigma^\dagger\sigma, H_{\rm NM}^{\rm 1e} ] = 0$, so populations do not change with time and the IBM Hamiltonian only describes pure dephasing of the emitter. Furthermore, the eigenstates of $H_{\rm NM}^{\rm 1e}$ are 
\begin{equation}
|\phi\rangle_g = |g\rangle\otimes|\{n_k\}\rangle,\; |\phi\rangle_e =|e\rangle\otimes B_-|\{n_k\}\rangle    
\end{equation}
with the associated eigenvalues 
\begin{equation}
E_g = \sum_k\omega_kn_k ,\;\quad E_e =\omega_0' + \sum_k\omega_kn_k,
\end{equation}
respectively. We defined $|g\rangle$ ($|e\rangle$) as the ground (excited) state of the emitter, $|\{n_k\}\rangle$ as the multimode phonon number state, $n_k$ as the number of phonons in the $k$-th mode, and $B_- = \Pi_k\mathfrak{D}(- g_k/\omega_k)$ with $\mathfrak{D}$ the displacement operator and $\omega_0' = \omega_0 -\int d\omega J_{\rm pn}(\omega)/\omega$ as the polaron-shifted frequency with $J_{\rm pn}(\omega)$ being the phonon spectral density.

To calculate the unitary operator $U(\tau)$, we diagonalize~\eqref{IBM} by using the unitary transformation $\tilde{H}_{\rm NM}^{\rm 1e} = \mbox{exp}(S) H_{\rm NM}^{\rm 1e} \mbox{exp}({-S})$, where $S = \sigma^\dagger\sigma\sum_k \left(g_k b_{k}^\dagger- g_k^*  b_{k}\right)/\omega_k$. This results in the diagonal Hamiltonian $\tilde{H}_{\rm NM}^{\rm 1e} = \omega_0'\sigma^\dagger\sigma + \sum_{k}\omega_{k} b_{k}^\dagger b_{k}$ with eigenstates $|g\rangle \otimes |\{n_k\}\rangle$ and $|e\rangle \otimes |\{n_k\}\rangle$ having eigenvalues $E_g$ and $E_e$ respectively.
Now we can define the unitary evolution as $U(\tau) = \mbox{exp}({-S})\mbox{exp}({-i\tilde{H}_{\rm NM}^{\rm 1e}t})\mbox{exp}(S)$. Writing $\mbox{exp}(S) = |g\rangle\langle g| +|e\rangle\langle e|B_+$, where $B_+ = \Pi_k\mathfrak{D}( g_k/\omega_k)$, 
we can express the evolution operator as 
\begin{equation}
U(\tau) = |g\rangle\langle g|U_{\rm P}(\tau) + \mbox{exp}(-i\omega_0'\tau)|e\rangle\langle e|B_-U_{\rm P}(\tau)B_+.
\end{equation}
Here $U_{\rm P}(\tau) = \mbox{exp}({-iH_{\rm pn}\tau})$, with $H_{\rm pn} \equiv \sum_{k}\omega_{k} b_{k}^\dagger b_{k}$~\cite{Nazir2016ModellingDots}. Therefore $\sigma(\tau) = \sigma B_-(\tau)B_+\mbox{exp}({-i\omega_0'\tau})$, and
\begin{eqnarray}
   g^{(1)}(\tau) &=& \gamma\mbox{Tr}\left[\sigma^\dagger\sigma B_-(\tau)B_+\mbox{e}^{-i\omega_0'\tau}\chi(0)\right]\nonumber = \gamma\langle \sigma^\dagger\sigma \rangle_{\rho(0)} \langle B_-(\tau)B_+ \rangle_{\rho_{\rm pn}}\mbox{e}^{-i\omega_0'\tau}=  \gamma\rho_{ee} \mathcal{C}(\tau)e^{-i\omega_0'\tau}.
   \label{g_1_1e_pn}
\end{eqnarray}
Here we have defined $\chi(0) = \rho_{\rm pn}\otimes\rho_{\rm pt}\otimes\rho(0)$ as the combined separable initial state of the system and the bath with $\rho_{\rm pn}$ the state of the phonon bath, $\rho_{\rm pt}$ the state of the photon bath, while $\rho$ is the system state.  Since there are no operators pertaining to the photon bath, it results in the trace over the photon bath state, which is unity. Thus one obtains the expression~\eqref{g_1_1e_pn} where $\rho_{ee} = \langle e |\rho(0) |e\rangle$, $\mathcal{C}(\tau)=\mathcal{C}_{\infty}(T) \mbox{exp}(\phi(\tau))$ is the phonon correlation function with $\mathcal{C}_\infty(T) = \mbox{exp}\left(-\int_0^\infty d\omega J_{\rm pn}(\omega)\coth(\beta\omega/2)/\omega^2\right)$ its steady-state value and $\phi(\tau) = \int_0^\infty d\omega J_{\rm pn} (\omega)(\cos(\omega\tau)\coth(\beta\omega/2)-i\sin(\omega\tau))/\omega^2$. 
Once the phonon bath has relaxed after causing a  fast dephasing of the emitter state on a picosecond scale, a steady state of phonon correlations is reached which can be associated with the formation of a polaron.
Then for $\tau>\tau_{\rm P}$ spontaneous emission and phonon-induced markovian dephasing due to the second-order emitter-phonon interactions take over, as described by the Hamiltonian 
\begin{equation}
H_{\rm MK}^{\rm 1e} = \omega_0\sigma^\dagger\sigma + \sum_qh_q(a_q\sigma^\dagger + a_q^\dagger\sigma) + \sigma^\dagger\sigma\sum_{k,k'} f_{k,k'} (b_{k}^\dagger + b_k)(b_{k'}^\dagger + b_{k'}) + \sum_{k}\omega_{k} b_{k}^\dagger b_{k},    
\end{equation}
    
 where the second term is the emitter-photon interaction term with $h_q$ the coupling between the emitter and the photon field of mode $q$, and $a_q$ ($a_q^\dagger$) the corresponding photon annihilation (creation) operator. The third term corresponds to the quadratic (second-order) emitter-phonon interaction term where $f_{k,k'}$ describes phonon-assisted virtual transitions to higher-lying states of the emitter that cause an inelastic scattering of phonons from mode $k$ to $k'$~\cite{Muljarov2004DephasingPhonons, Reigue2017ProbingDots, Denning2020PhononSources}.  We consider spontaneous decay and pure dephasing due to the second-order coupling as Markovian processes that are described by the Markovian master equation~\cite{Reigue2017ProbingDots}, 
\begin{equation}
    \frac{d\rho_s}{dt}= \gamma(\sigma\rho\sigma^\dagger-1/2\{\sigma^\dagger\sigma,\rho_s\}_+) + 2\gamma_{\rm pd}(T)(\sigma^\dagger\sigma\rho\sigma^\dagger\sigma-\rho_s).
    \label{ME_1e}
\end{equation}  
To calculate the two-time expectation value of the field operators, i.e. $g^{(1)}(\tau > \tau_{\rm P}) = \gamma\langle \sigma^\dagger\sigma(\tau)\rangle_{\rm ME}$, we will use the quantum regression theorem (QRT), which allows us to use the markovian master equation (ME),~\eqref{ME_1e} to calculate the two-time expectation values~\cite{Carmichael2000StatisticalEquations} which are given by $g^{(1)}(\tau > \tau_{\rm P}) = \gamma \text{exp}(-\Gamma \tau) $, where $\Gamma = \gamma/2 + \gamma_{\rm pd}(T)$. Therefore, by concatenating the dynamics of the two time intervals, we obtain the first-order correlation function at all times as
    \begin{equation}
    g^{(1)}(\tau) = \gamma\rho_{ee}\mathcal{C}(\tau)\mbox{e}^{-(\Gamma + i\omega_0')\tau}.\;\;\;
     \label{g_1_1e_app}
\end{equation}
Now, if we are only interested in the zero-phonon line (ZPL) of the spectrum and not so much in the sidebands, then in the time domain we can make the simplification to set $\mbox{exp}(\phi(\tau)) = 1$~\cite{Rouse2019OptimalEnvironment}, which results in $\mathcal{C}(\tau) \rightarrow \mathcal{C}_\infty(T)$. Using this in \eqref{g_1_1e_app} we obtain
    \begin{equation}
    g^{(1)}(\tau) = \gamma\rho_{ee}\mathcal{C}_\infty(T)\mbox{e}^{-(\Gamma + i\omega_0')\tau}.\;\;\;
     \label{g_1_1e_IS}
\end{equation}
\eqref{g_1_1e_IS} is a consequence of discarding the fast initial non-markovian dynamics of the phonon bath but keeping track of the steady-state value of the phonon correlations $\mathcal{C}_\infty(T)$, which is then utilized as an initial condition to the markovian master equation. Such a scheme is referred to as the initial-slip scheme~\cite{Suarez1992MemorySystems, Yu2000Post-markovSystems}.
\section{Derivation of $g^{(1)}(\tau)$ from concatenation scheme for $N$ emitters}
\label{2E_exact}

We start with the Hamiltonian where $N$ identical emitters are coupled to a phonon bath,
\begin{eqnarray}
 H_{\rm NM}^{\rm Ne} = \sum_{n=1}^N\left(\omega_0\sigma^\dagger_n\sigma_n + \sigma^\dagger_n\sigma_n\sum_k\left(g_k^{(n)} b_{k}^\dagger +g_k^{*(n)}  b_{k}\right) \right) + \sum_{k}\omega_{k} b_{k}^\dagger b_{k}.
\end{eqnarray}
We apply the polaron transformation $\tilde{H}_{\rm NM}^{\rm Ne} = \mbox{exp}(S) H_{\rm NM}^{\rm Ne} \mbox{exp}(-S)$ where $\mbox{exp}(S) = \mbox{exp}(\sum_{n = 1}^N S_n)$, and $S_n = (g_k^{(n)}b^\dagger - g_k^{*(n)}b)/\omega_k$, to the above Hamiltonian. We work in the single-excitation subspace with $N$ orthogonal states  $|n\rangle \equiv |g, g,...,e_n,...\rangle$ corresponding to the excitation of only the $n^{\rm th}$ emitter.

This helps to rewrite the polaron transformation as $\mbox{exp}(S)    = |0\rangle \langle 0 |+\sum_{n=1}^N|n\rangle \langle n |B_+^{(n)}$, where $|0\rangle = |g, g,...,g,...\rangle$ is the collective ground state. Applying this unitary transformation to the Hamiltonian, we can write the above Hamiltonian into the diagonal form 
\begin{equation}
    \tilde{H}_{\rm NM}^{\rm Ne} = \sum_{n=1}^N\omega_0'|n\rangle \langle n |+ \sum_{k}\omega_{k} b_{k}^\dagger b_{k}.
    \label{polaron_2em}
\end{equation}
The unitary time evolution operator $U(\tau) \equiv \mbox{exp}(-i{H}_{\rm NM}^{\rm Ne}\tau)$ can then be written as $U(\tau) = \mbox{exp}({-S})\mbox{exp}({-i\tilde {H}_{\rm NM}^{\rm Ne}\tau})\mbox{exp}(S)$, and takes the  form
\begin{equation}
    U(\tau) = |0\rangle \langle 0|U_{\rm P} + \sum_{n=1}^N|n\rangle \langle n|\mathcal{B}_n,
    \label{U_2e_app}
\end{equation}
where $\mathcal{B}_n = \mbox{exp}({-i\omega_0' \tau}) B_-^{(n)}U_{\rm P} B_+^{(n)}$ and $B_\pm^{(n)} = \Pi_k\mathfrak{D}(\pm g_k^{(n)}/\omega_k)$, with $g_k^{(n)} \equiv |g_k|\mbox{exp}(i\mathbf{k}\cdot \mathbf{r}_n)$.  Using \eqref{U_2e_app}, we calculate the time evolution of the atomic operators as $\sigma_n(\tau) = \sigma_nU_{\rm P}^\dagger(\tau)\mathcal{B}_n$. Therefore,
\begin{eqnarray}
     g^{(1)}(\tau) &=& \sum_{n,m =1}^N\gamma_{nm}\mbox{Tr}\left[\sigma_n^\dagger\sigma_mU^\dagger_{\rm P}(\tau)B_-^{(n)}U_{\rm P} B_+^{(n)}\mbox{e}^{-i\omega_0'\tau}\chi(0)\right]\nonumber= \sum_{n,m =1}^N\gamma_{nm}\langle \sigma_n^\dagger\sigma_m \rangle_{\rho(0)} \langle B_-^{(n)}(\tau) B_+^{(n)} \rangle_{\rho_{\rm pn}} \mbox{e}^{-i\omega_0'\tau}.\nonumber\\
    & = &  \mathcal{C}(\tau)\sum_{n,m=1}^N\gamma_{nm}\rho_{nm}e^{-i\omega_0' \tau}.
\end{eqnarray}
Here, as in the supplementary Sec.~\ref{secii}, we have defined $\chi(0) = \rho_{\rm \rm pn}\otimes\rho_{\rm pt}\otimes\rho(0)$. Also, $\rho_{nm} = \langle n | \rho(0) | m\rangle$, and $\mathcal{C}(\tau)=\mathcal{C}_\infty(T) \mbox{exp}({\phi(\tau)})$ is the phonon correlation function. Notice that in the derivation of the above equation the phonon cross correlations do not appear naturally.

Next, we deduce the decay dynamics for $\tau>\tau_{\rm P}$ via the following multi-emitter Markovian master equation derived in the nearest-neighbour interaction approximation that considers only the dipole-dipole interaction $\omega_{nm}$ of its nearest neighbour i.e.  $\omega_{nm} = \omega_{\rm col}$ if $|n - m| = 1$ and zero otherwise where $n$ and $m$ are the $n$-th and $m$-th emitters respectively~\cite{Reitz2022CooperativePlatforms}. This gives the master equation
\begin{multline}
   \frac{d\rho}{dt} = -i\sum_{n=1}^N\left[\omega_{\rm col}\left(\sigma_n^\dagger\sigma_{n + 1} + \sigma_{n + 1}^\dagger\sigma_{n}\right), \rho\right]  +  \sum_{n,m=1}^N\gamma_{nm}\left(\sigma_n\rho\sigma_{m}^\dagger-\frac{1}{2}\left\{\sigma_n^\dagger\sigma_{m},\rho\right\}_+\right)\\+ 2\gamma_{\rm pd}(T)\sum_{n=1}^N\left(\sigma_n^\dagger\sigma_{n}\rho\sigma_n^\dagger\sigma_{n}-\frac{1}{2}\left\{\sigma_n^\dagger\sigma_{n},\rho\right\}_+\right),
   \label{me_Ne_0}
\end{multline}
where we have also added the dissipator that causes markovian pure dephasing due to the local phonon baths. The phonon correlation due to an excited emitter become uncorrelated at $\tau > \tau_{\rm P}$ since $\mathcal{C}(\tau) \rightarrow \mathcal{C}_\infty(T)$, giving the phonon correlation length $L_{\rm P} = c\tau_{\rm P}$, where $c$ is the speed of sound. Since we work in the single excitation regime and we choose the inter-emitter separation $r_{12} > L_{\rm P}$, the phonon field already decays before reaching the adjacent emitter. Therefore we can assume uncorrelated inter-emitter phonon correlations implying independent phonon baths. The first-order correlation function $g^{(1)}(\tau) = \sum_{n,m = 1}^N\gamma_{nm}\langle \sigma^\dagger_n\sigma_m(\tau)\rangle_{\rm ME}$ can be calculated from the QRT using the above master equation. In deriving \eqref{me_Ne_0}, we have considered only the emission of a photon characterized by the collapse operator $\sigma_n$ and neglected absorption of a photon in the environment characterized by the collapse operator $\sigma_n^\dagger$. For a 1D chain of emitters, this equation can be diagonalised (approximately) by the prescription given in Ref.~\cite{Reitz2022CooperativePlatforms}, which we will use and briefly outline. Expressing the operators $\sigma_n$ in terms of the collective operators via the relation $\sigma_n = \sum_{l=1}^NM_{nl}\sigma_l$ in \eqref{me_Ne_0} exactly diagonalizes the coherent part of the Hamiltonian. On the other hand, applying the same transformation to the dissipative part results in off-diagonal terms of the form $\sigma_l^\dagger\sigma_{l'}$. This implies that the simultaneous diagonalization of the coherent and dissipative part of the master equation for $N > 2$ is generally not possible. We can still find a simultaneous diagonalization of the coherent and dissipative parts by accepting only the diagonal terms in the dissipative part of the transformed equation by the approximation $l = l'$. This approximation results in the  diagonal form of the equation describing $N$ emitters in a 1D chain with only the nearest-neighbour interactions $\omega_{\rm col}$ taken into account, 
\begin{equation}
    \frac{d\rho}{dt} = \sum_{l=1}^N\left(-i\left[\omega_l\sigma_l^\dagger\sigma_l,\rho\right] +  \gamma_l\mathcal{D}_{\sigma_l}(\rho)  + 2\gamma_{\rm pd}(T)\mathcal{D}_{\sigma_l^\dagger\sigma_l}(\rho)\right),
    \label{dme_Ne}
\end{equation}
where $\omega_l = 2\omega_{\rm col}\cos\left(\pi l/(N +1)\right)$, $\sigma_l = \sum_n M_{nl}|\sigma_n\rangle$, and $\gamma_l = \sum_{nm}\gamma_{nm} M_{nl}M_{ml}$  with $M_{nl}\equiv \sqrt{2/(N + 1)}\sin(\pi n l/ (N + 1))$~\cite{Plankensteiner2015, Reitz2022CooperativePlatforms} and $\mathcal{D}_{A}(\rho)\equiv A \rho A^\dagger-\frac{1}{2}\{A^\dagger A,\rho\}_+ $, with $\{\}_+$ the anticommutator.

Utilizing this same approximate diagonalization scheme, we find the first-order correlation function for $\tau>\tau_{\rm P}$, using the QRT in terms of the collective states as $g^{(1)}(\tau) = \sum_{l= 1}^N\gamma_{l} \langle \sigma_{l}^\dagger\sigma_{l}(\tau)\rangle_{\rm DME}$, employing the diagonal master equation (DME) in \eqref{dme_Ne}. Thus, upon concatenating the dynamics, we obtain the first-order correlation function at all times in the collective basis as
\begin{equation}
 g^{(1)}(\tau) = \mathcal{C}(\tau)\sum_{l =1}^N \gamma_l\rho_{ll} \mbox{e}^{-(\Gamma_l + i\omega_l' )\tau},\;\;\;\forall\tau,    
  \label{2e_corr_tot_app}
\end{equation}
where $\omega_l' = \omega_0' + 2\omega_{\rm col}\cos(\pi l/(N +1)) $. Now using as before the ''initial-slip'' simplification $\mathcal{C}(\tau)\rightarrow\mathcal{C}_{\infty}(T)$, we obtain simplified analytical expressions for the correlation function
\begin{equation}
 g^{(1)}(\tau) = \mathcal{C}_\infty(T)\sum_{l =1}^N \gamma_l\rho_{ll} \mbox{e}^{-(\Gamma_l + i\omega_l' )\tau},\;\;\;\forall\tau.    
  \label{2e_corr_tot_IS}
\end{equation}
Its Fourier transform gives the collective-emission spectrum near the ZPL, which is accurate around the ZPL.
\section{Derivation of $g^{(1)}(\tau)$ for $N$ emitters from the polaron master equation}
The dynamics of $N$ identical quantum emitters interacting with both phonons and photons in a bulk environment can be described by the  Hamiltonian
\begin{equation}
    H = \sum_{n=1}^N\left(\omega_0\sigma^\dagger_n\sigma_n +\sum_j\left(h_j^{(n)}a_j\sigma^\dagger_i+h_j^{*(n)}a_j^\dagger\sigma_i\right) + \sigma^\dagger_n\sigma_n\sum_k\left(g_k^{(n)} b_{k}^\dagger + g_k^{*(n)}  b_{k}\right) \right)\\  + \sum_j\omega_ja_j^\dagger a_j + \sum_{k}\omega_{k} b_{k}^\dagger b_{k},
    \end{equation}
where $h_j^{(n)} \equiv |h_j|\rm \mbox{exp}({i\mathbf{q}\cdot \mathbf{r}_{n}})$ and $g_k^{(n)}\equiv |g_k|\rm \mbox{exp}({i \mathbf{k}\cdot \mathbf{r}_{n}})$ are the position-dependent emitter-photon and emitter-phonon couplings, respectively. We apply the unitary polaron transformation $\tilde{H} = \mbox{exp}(S) H \mbox{exp}({-S})$ to the above Hamiltonian  where $S$  was defined in the supplementary Sec.~\ref{2E_exact} and obtain
\begin{equation}
    \tilde H = \sum_{n=1}^N\left(\omega_0'\sigma^\dagger_n\sigma_n +\sum_j\left(h_j^{(n)}a_j\sigma^\dagger_n B_+^{(n)}+h_j^{*(n)}a_j^\dagger\sigma_nB_-^{(n)}\right) \right) + \sum_j\omega_ja_j^\dagger a_j + \sum_{k}\omega_{k} b_{k}^\dagger b_{k}.
    \label{Polaron_ME2}
\end{equation}
From this equation, we can see that the polaron transformation dresses the system operators with the phonon bath operators and leads to a polaron-shifted frequency $\omega_0'$ of the emitters as described in the supplementary Sec.~\ref{secii}. This allows us to derive a polaron master equation with the interaction Hamiltonian $\tilde H_I = \sum_{n,j}\left(h_j^{(n)}a_j\sigma^\dagger_n B_+^{(n)}+h_j^{*(n)}a_j^\dagger\sigma_nB_-^{(n)}\right) $ as the perturbation with the nearest-neighbour interaction 
\begin{equation}
 \frac{d\rho}{dt}= -i\sum_{n=1}^N\bigg[\omega_0'  \sigma_n^\dagger\sigma_n + \Omega_{\rm col}\bigg(\sigma_n^\dagger\sigma_{n + 1} + \sigma_{n + 1}^\dagger\sigma_{n}\bigg), \rho\bigg]  +  \sum_{n,m= 1}^N \Upsilon_{nm}\mathcal{D}_{\sigma_{n,m}} + 2\gamma_{\rm pd}(T)\sum_{n=1}^N \mathcal{D}_{\sigma_{n}^\dagger\sigma_n},   
 \label{SPME_nd_app}
\end{equation}
where in the final term, we have also added the pure dephasing due to the second-order phonon coupling, which has been explicitly evaluated in Ref. \cite{Reigue2017ProbingDots}.  The decay constants in the above equation calculated from the bath correlations functions have both photonic and phononic characters. The master equation takes the standard form of a coherent evolution (in square brackets) and a non-unitary evolution due to the decay in the system characterised by the dissipators $\mathcal{D}_{\sigma_{n,m}} \equiv  \sigma_n\rho \sigma_m^\dagger-\frac{1}{2}\{\sigma_m^\dagger \sigma_n,\rho\}$ and $\mathcal{D}_{\sigma_{n}^\dagger\sigma_n} \equiv  {\sigma_{n}^\dagger\sigma_n}\rho {\sigma_{n}^\dagger\sigma_n}-\frac{1}{2}\{\sigma_n^\dagger \sigma_n,\rho\}$ with inter-emitter decay rates given by $\Upsilon_{nm}$ (decay) and $2\gamma_{\rm pd}(T)$ (pure dephasing) respectively. Due to the phonon interactions, the dipole-dipole interaction $\Omega_{\rm col}$ and inter-emitter decay rates $\Upsilon_{nm}$ get renormalized, which will be discussed in the next section. \newline
The first-order correlation function can be calculated by using the polaron-transformed Hamiltonian~\eqref{Polaron_ME2}, using the Heisenberg equation of motion, giving
\begin{equation}
g^{(1)}(\tau) = \sum_{n,m = 1}^N\mathcal{C}_{nm}(\tau)\gamma_{nm}\langle \sigma_n^\dagger\sigma_m(\tau)\rangle_{\rm PE},
\label{corr_pm}
\end{equation}
where the subscript PE means that we need to evaluate the correlation using the master equation in the polaron frame given by \eqref{SPME_nd_app}, while $\mathcal{C}_{nm}(\tau) = \langle B_-^{(n)}(\tau) B_+^{(m)} \rangle $, where  $\mathcal{C}_{nm}(\tau) = \mathcal{C}(\tau)$ for $n = m$ while $\mathcal{C}_{nm}(\tau) = \mathcal{C}_\infty(T)$ for $n \neq m$, the latter being a consequence of the assumption that the phonon baths at different emitter positions are uncorrelated~\cite{Rouse2019OptimalEnvironment}. 

We can yet again use the collective basis and the diagonalization scheme explained in the previous section to rewrite \eqref{corr_pm} in the collective basis upon suppressing the sidebands as $g^{(1)}(\tau) = \mathcal{C}_{\infty}(T)\sum_{l}\gamma_{l}\langle \sigma_l^\dagger\sigma_l(\tau)\rangle_{\rm DPE}$ where now the correlation is evaluated using the diagonal form of multi-emitter polaron master equation
\begin{equation}
    \frac{d\rho}{dt} = \sum_{l=1}^N\left(-i\left[\Omega_l'\sigma_l^\dagger\sigma_l,\rho\right] +  \Upsilon_l\mathcal{D}_{\sigma_l}(\rho)  + 2\gamma_{\rm pd}(T)\mathcal{D}_{\sigma_l^\dagger\sigma_l}(\rho)\right),
    \label{pme_Ne}
\end{equation}
where $\Upsilon_l = \sum_{nm}\Upsilon_{nm} M_{nl}M_{ml}$, and $\Omega_l' = \omega_0' + 2\Omega_{\rm col}\cos(\pi l/(N +1))$.  Thus, the first-order correlation function becomes
\begin{equation}
g^{(1)}(\tau) = \mathcal{C}_\infty(T) \sum_{l =1}^N \gamma_l\rho_{ll} \mbox{e}^{-(\digamma_l + i\Omega_l' )\tau},\;\;\;\forall\tau ,   
\label{2e_corr_tot_app_2}
\end{equation}
where $\digamma_l = \Upsilon_l/2 + \gamma_{\rm pd}(T)$. In the next section we will see how these collective decay rates $\Upsilon_{nm}$ and the collective Lamb shift $\Omega_{\rm col}$ obtained from the polaron master equation are modified due to the phonons.

\section{Derivation of phonon-renormalized decay rates in polaron method}
\label{Phonon_dressed_emission}
The bath correlations of the emitters coupled to photon and phonon baths obtained from the multi-emitter polar master equation~\eqref{SPME_nd_app} derived in the previous section have the form
\begin{equation}
 \Gamma_{nm}(\tau) = \langle G_n^\dagger(\tau) G_m B_+^{(n)}(\tau)B_-^{(m)}\rangle_{\rho_{\rm pt} \otimes \rho_{\rm pn}} = \langle G_n^\dagger(\tau) G_m \rangle_{\rho_{\rm pt}}\langle B_+^{(n)}(\tau)B_-^{(m)}\rangle_{\rho_{\rm pn}},
 \label{2bath_corr_app}
\end{equation}
where $G_n(\tau) = \sum_q h_q^{*(n)}a_q^\dagger  \mbox{exp}({i\omega_q \tau}) + \rm h.c.$, is a linear combination of photon operators. We assume the associated photon density matrix $\rho_{\rm pt}$  to be the vacuum state. Similarly, $B_+^{(n)}(\tau) = \Pi_k \mathfrak{D}({g_k^{(n)}\mbox{exp}({i\omega \tau}})/{\omega_k})$ is a function of phonon operators, and the initial phonon density matrix $\rho_{\rm pn}$ is assumed to be a thermal state and $\mathfrak{D}$ is the displacement operator. The phonon bath correlations in the assumption of independent phonon baths can be given by $\langle B_+^{(n)}(\tau)B_-^{(m)}\rangle = \mathcal{C}_\infty(T) \mbox{exp}({\phi(\tau)\delta_{nm}})$, where $\delta_{nm}$ is the Kroneker delta function. On the other hand, the photon correlations are
\begin{equation}
    \langle G_n^\dagger(\tau) G_m \rangle_{\rho_{\rm pt}} =\sum_q |h_q|^2 \mbox{e}^{i\big(\textbf{q}\cdot\mathbf{(r_n-r_m)}-\omega_q\tau\big)},
    \label{G_1}
\end{equation}
where $h_{q}  = (\omega_q/ 2\mathcal{V}\epsilon_0)(\mathbf{d}\cdot \hat{\varepsilon}_q)$, is the coupling constant where $\mathcal{V}$ is the mode volume, $\epsilon_0$ is the permittivity of free space, $\mathbf{d}$ is the transition dipole moment and $\hat{\varepsilon}_q$ is the unit vector associated to the transverse components of the electromagnetic waves. The mode index $q$ labels both the wavevector and corresponding polarization indices of transverse plane waves.  Furthermore, $\mathbf{r}_i$ is the position of $i$-th emitter and $\mathbf{q} \equiv \mbox{n}\mathbf{q}_0$, with $\rm n$ being the refractive index of the medium and $\mathbf{q}_0$ is the free space wavevector. In arriving at \eqref{G_1} we have also used $\langle a_qa_q^\dagger\rangle_{\rho_{\rm pt}} = 1$, since the photon number $n(\omega_q) = \langle a_q^\dagger a_q\rangle_{\rho_{\rm pt}} = 0 $ for the photon state assumed to be in a vacuum state.

\subsection{Single-emitter decay rate}
\label{2bath_corr_app2}
When $n = m$ we obtain from \eqref{G_1} that $\langle G_n^\dagger(\tau) G_n \rangle_{\rho_{\rm pt}} =\sum_q |h_q|^2 \mbox{exp}({-i\omega_q\tau})$. Using the definition of the photon spectral density  $J_{\rm pt} (\omega) = \sum_q |h_q|^2\delta(\omega-\omega_q)$, we get $\langle G_n^\dagger(\tau) G_n \rangle_{\rho_{\rm pt}} = \int_0^\infty d\omega J_{\rm pt}(\omega) \mbox{exp}({-i\omega\tau})$. Using this definition in \eqref{2bath_corr_app}, we can find the frequency-dependent correlation function as $\Gamma_{nn}(\nu)=\int_0^\infty d\tau \mbox{exp}({i\nu\tau})\Gamma_{nn}(\tau)$. After rearranging, this becomes
\begin{equation}
 \Gamma_{nn}(\nu)  = \int_0^\infty d\omega \, J_{\rm pt}(\omega) \mathcal{K}_{\rm pn}(\nu - \omega),
 \label{Gamma_nn}
\end{equation}
where
\begin{equation}
\mathcal{K}_{\rm pn}(\xi) = \int_0^\infty d\tau \, \mathcal{C}(\tau) \mbox{e}^{i\xi\tau}
\end{equation}
is the kernel describing the phonon interaction.
The above equation demonstrates that the photon and phonon interactions are convolved, exhibiting non-additivity of the two baths~\cite{Rouse2022AnalyticFrame}. An analytical expression of the above expression are hard to obtain and require further approximations~\cite{Nazir2016ModellingDots, Rouse2022AnalyticFrame}. 
Since most of the light emission happens around the resonance frequency, it can be a good approximation to assume a frequency-independent photon spectral density, which constitutes what is referred to as a flat spectral density approximation. For our nondispersive bulk medium, this implies  $J_{\rm pt}(\omega) \approx \eta \propto \omega^3$, which allows us to simplify~\eqref{Gamma_nn} as~\cite{Nazir2016ModellingDots, Rouse2022AnalyticFrame},
\begin{eqnarray}
\Gamma_{nn}(\nu) &=& \int_0^\infty d\tau \, \mathcal{C}(\tau)\int_0^\infty d\omega J_{\rm pt}(\omega)\mbox{e}^{i(\nu - \omega)\tau} = \eta \int_0^\infty d\tau \, \mathcal{C}(\tau)\mbox{e}^{i\nu\tau}\left(\pi\delta(\tau)-i\mathcal{P}\left(\frac{1}{\tau}\right)\right).
\end{eqnarray}
Using the relation $\mathcal{C}(0) = 1$ and evaluating the self-induced interaction $\Gamma_{nn}$ in the polaron master equation \eqref{SPME_nd_app} we substitute $\nu = \omega_0'$. We thus find the decay rate as the real part of the $\Gamma_{nn}$ which is given as $\gamma(\omega_0') \approx 2\pi\eta \propto \omega_0'^3$. Therefore,  within the flat spectral density approximation, the net effect of transitions involving all possible phonon states is identical to the single electronic transition between the phonon-renormalized excited state $\omega_0'$ and the ground state. 

\subsection{Inter-emitter decay rates}
Since the phonon baths for the individual emitters are assumed to be independent, they do not have a spatial correlation, which implies $\langle B_+^{(n)}(\tau)B_-^{(m)}\rangle_{\rho_{\rm pn}} = \mathcal{C}_\infty(T)$ for $n \neq m$ \cite{Rouse2019OptimalEnvironment}. Therefore from~\eqref{2bath_corr_app} and~\eqref{G_1} the total correlation becomes
\begin{equation}
    \Gamma_{nm}(\tau) =  \sum_q |h_q|^2 \mbox{e}^{i\big(\textbf{q}\cdot\mathbf{(r_n-r_m)}-\omega_q\tau\big)}\mathcal{C}_\infty(T). 
\end{equation}
Since the phonon correlation is  a temperature-dependent but $q$-independent parameter, it can be taken out of the summation. Because of this, the assumption of independent phonon baths for the various emitters will allow us to derive simple analytical results for the inter-emitter emission rates, without the need to invoke a flat spectral density (or other) approximation for the photon baths. (Recall that for the single-emitter emission rate in Sec.~\ref{2bath_corr_app2}, we did make this assumption.) So here, essentially, we are left with computing the inter-emitter decay rates associated with the photon bath alone. Such computations have already been performed in many places, for example, in Refs.~\cite{Lehmberg1970RadiationFormalism, ficek2005quantum}. For completeness, we give a brief sketch of the derivation for bulk media. The sum is converted into an integral by using the transformation $\sum_q\rightarrow\mathcal{V}/(2\pi)^3\int d^3q$, where $\mathcal{V}$ is the mode volume. With some algebra, we obtain
\begin{equation}
\Gamma_{nm}(\tau) = \frac{\mathcal{C}_\infty(T)|d|^2}{4\pi^2c^3\epsilon_0}\int_0^\infty d\omega_q \, \omega_q^3\left(\mathbb{I} + \frac{\hat{d}\cdot \nabla_R}{q^2}\right)\frac{\sin q R}{q R} \mbox{e}^{-i\omega_q\tau},
\label{Gamma_nm_app}
\end{equation}
where $R = |r_n - r_m|$. Subsequently, one obtains the optical inter-emitter interactions $\Gamma_{nm}(\omega)$ by evaluating the Fourier transform (FT) of \eqref{Gamma_nm_app} where the FT of the term inside the integral has a known form~\cite{Lehmberg1970RadiationFormalism, ficek2005quantum}. Thus, one can write the complex optical inter-emitter interactions as  
\begin{equation}
 \Gamma_{nm}(\omega)= \left(\frac{\gamma_{nm}(\omega)}{2} + i \omega_{nm}(\omega)\right)\mathcal{C}_\infty(T) = \frac{\Upsilon_{nm}(\omega)}{2} + i \Omega_{nm}(\omega)\;\;\; \mbox{for}\; {n\neq m},
 \label{gamma_phonons}
\end{equation}
where $\Upsilon_{nm}(\omega)$ is the phonon-renormalized inter-emitter decay rate while $\Omega_{nm}(\omega)$ is the phonon-renormalised dipole-dipole interaction which needs to be evaluated at $\omega = \omega_0'$ in the polaron master equation~\eqref{SPME_nd_app}. Both for the $\Upsilon_{nm}(\omega)$ and the $\Omega_{nm}(\omega)$, the phonon renormalization amounts to multiplication by the factor $\mathcal{C}_\infty(T)$. In the nearest-neighbor approximation, $\Omega_{nm}(\omega)$ equals $\Omega_{\rm col}(\omega)$ for nearest neighbors and vanishes otherwise.
\section{Simulations at emitter separation $\lambda/15$}
In this section, we plot the emission spectrum for two emitters separated by a distance of $\lambda/15$ interacting via dipole-dipole interaction in the presence of phonon baths in Fig.~\ref{fig:2e_15} similar to Fig.~1 in the main text where we used an inter-emitter separation of $\lambda/25$. We also plot the normalized integrated intensity for emitters separated by $r_{12} = \lambda/25$ in Fig.~\ref{fig:biexpo}(a) and for $r_{12} =  \lambda/15$ in Fig.~\ref{fig:biexpo}(b) using  
\begin{equation}
\langle E^{-}(\tau)E^{+}(\tau)\rangle = \langle \sigma^\dagger_+\sigma_+\rangle \mbox{exp}(-\gamma_+\tau) + \langle \sigma^\dagger_-\sigma_-\rangle \mbox{exp}(-\gamma_-\tau),
\label{nII}
\end{equation}
where $\langle \sigma_\pm^\dagger\sigma_\pm\rangle = \langle \pm|\rho(0)|\pm\rangle$. We see the bi-exponential behaviour of the emission due to the initial two-emitter state $|\psi(0)\rangle = |e,g\rangle = (|+\rangle + |-\rangle)/\sqrt{2}$, where $|+\rangle$ ($|-\rangle$) is the superradiant (subradiant) state.   
\setcounter{figure}{0}   
\begin{figure}[h]
    \centering
    \includegraphics[width = 0.5\textwidth]{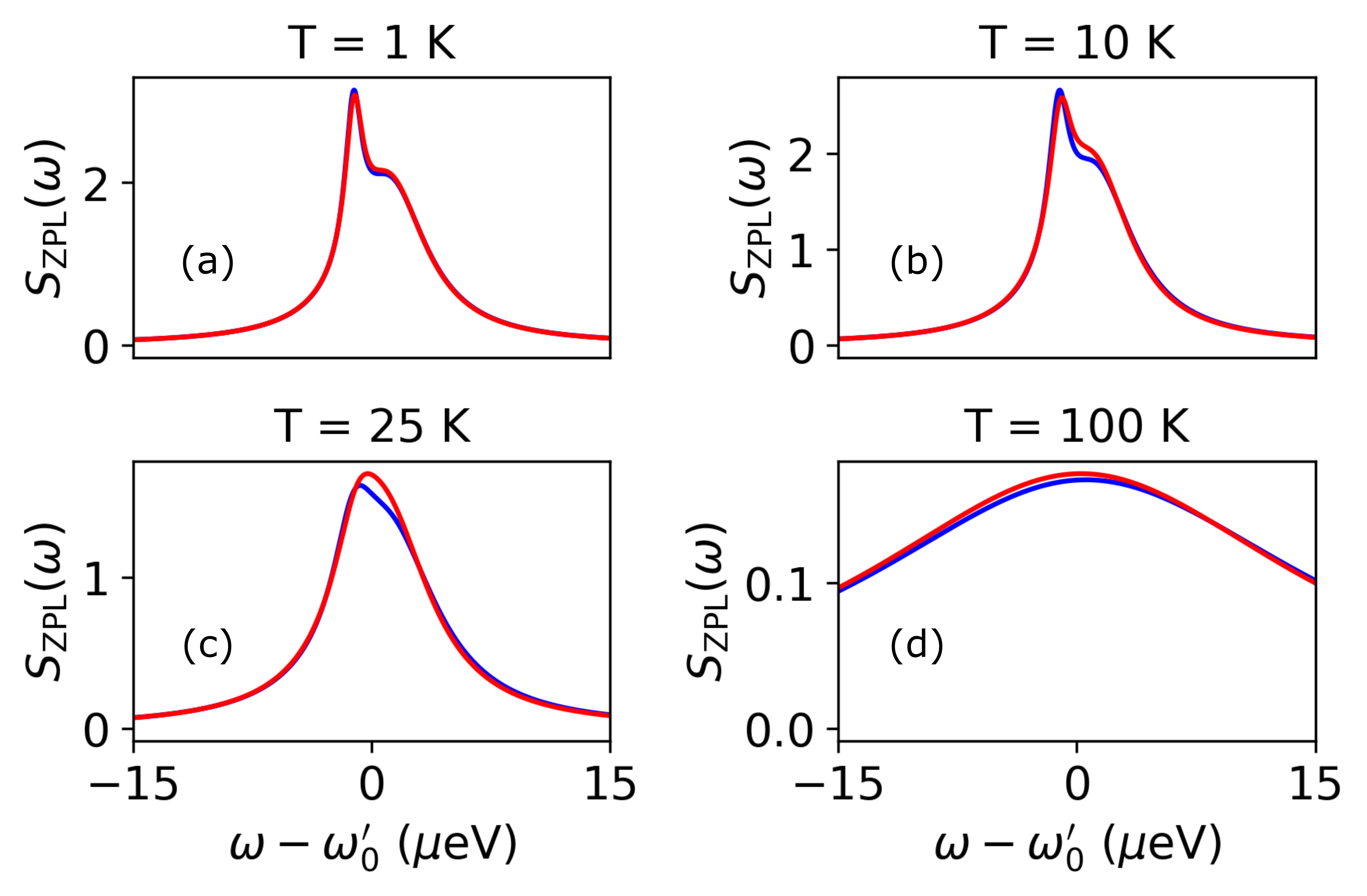}
    \caption{Temperature-dependent emission spectra for two emitters prepared in an initial state $|\psi(0)\rangle = |e,g\rangle$, now separated by a distance of $\lambda/15$, calculated using the concatenation approach (in blue) and the polaron approach (in red). Otherwise we use the same simulation parameters as in Fig.~\ref{fig1:1ecorr}.}
    \label{fig:2e_15}
\end{figure}

\begin{figure}[h]
    \centering
\begin{subfigure}
    \centering
        \includegraphics[width = 0.25\textwidth]{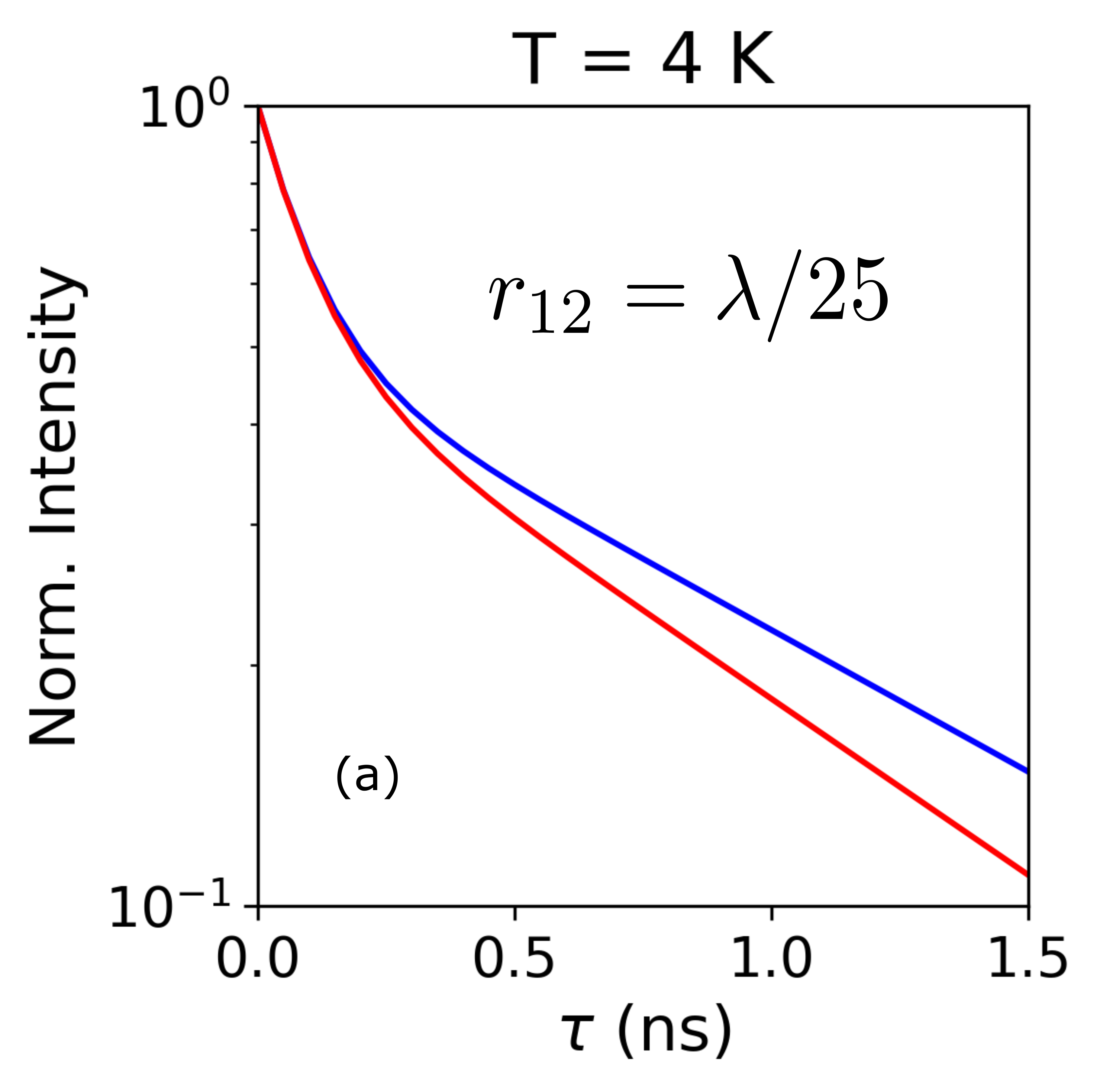}
\end{subfigure}
\begin{subfigure}
    \centering
        \includegraphics[width = 0.25\textwidth]{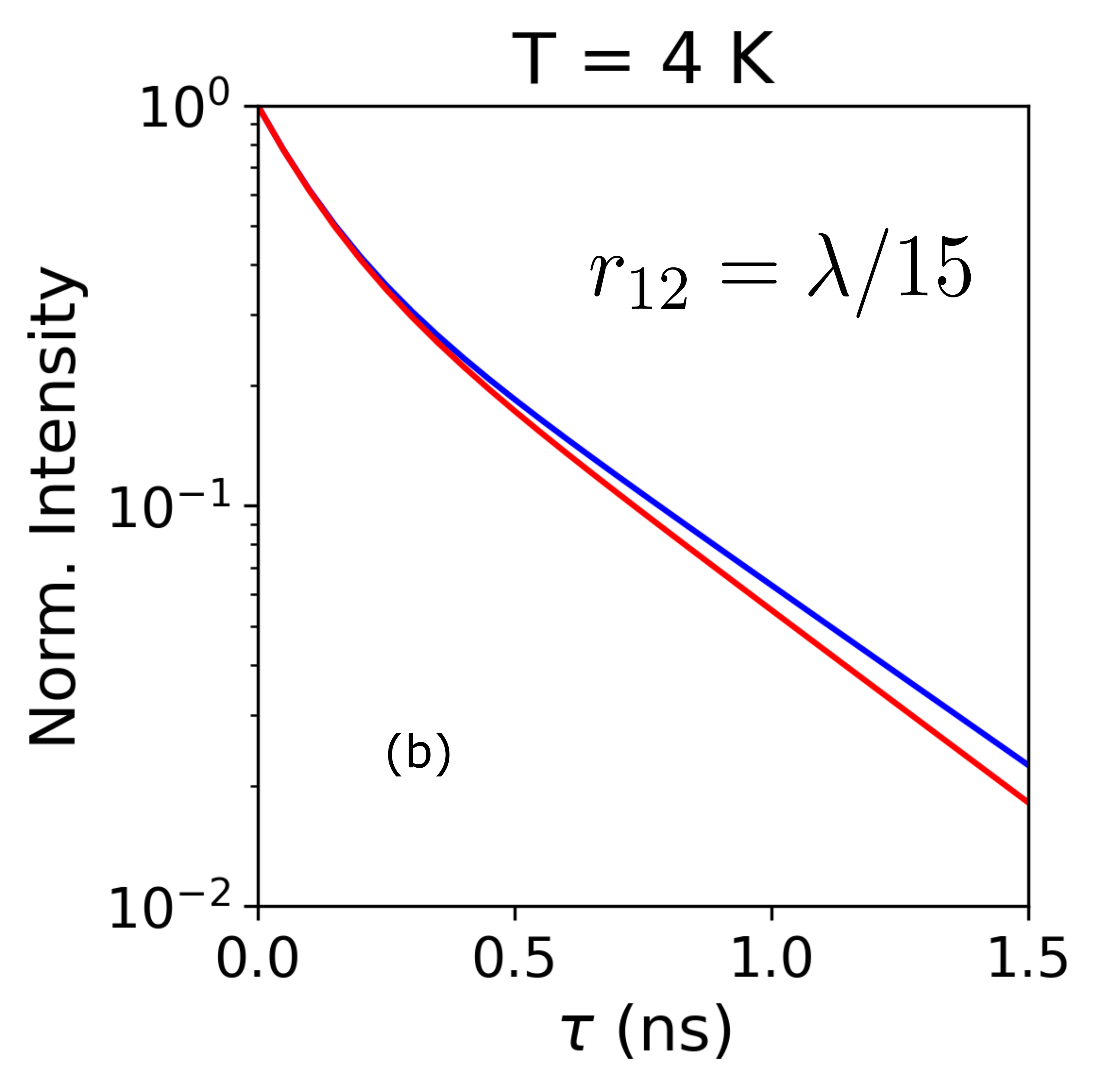}
\end{subfigure}
    \caption{Bi-exponential dynamics of the normalized integrated intensity of two interacting emitters prepared in an initial state $|\psi(0)\rangle = |e,g\rangle$, based on \eqref{nII}, calculated in the concatenation scheme (in blue solid lines) and with the polaron method (in red solid lines). In panel (a) the separation between the emitters is $\lambda/25$ exhibiting decay rate enhancement $\Gamma_+/\Gamma_- = 13.1\approx \digamma_+/\digamma_- $, with $\Gamma_+ = 1.58$~GHz and $\Gamma_- = 0.12$~GHz. In panel~(b), the separation between emitters is $\lambda/15$, which results in  $\Gamma_+/\Gamma_- = 4.45\approx \digamma_+/\digamma_- $, with $\Gamma_+ = 1.38$~GHz and $\Gamma_- = 0.31$~GHz.} 
    \label{fig:biexpo}
\end{figure}

\section{Derivation of the $N$-emitter initial system state at $\tau = \tau_{\rm P}$ in concatenation scheme}
In our concatenation scheme the time evolution of the system state for $\tau \leq \tau_{\rm P}$ is given by $\rho(\tau) = \mbox{Tr}_{\rm E}\left[U(\tau) \chi(0) U^\dagger(\tau)\right]$ where $\chi(0)$ is the combined direct-product state of the system and the environment defined in previous sections, and $U(\tau)$ is the unitary operator given by \eqref{U_2e_app}. In the resulting reduced system density matrix, apart from the diagonal matrix elements corresponding to the basis vectors, $|0\rangle$ and other states $|n\rangle$ constituting the single-excitation manifold (SEM), only those off-diagonal elements that contribute to the transitions within these SEM states are allowed, which leads to the time evolution of the system state for ($\tau \leq \tau_{\rm P}$)
\begin{eqnarray}
\rho(\tau) &=& \sum_{n = 0}^N|n\rangle\langle n|\rho_{nn} + \sum_{n \neq m = 1}^N |n\rangle\langle m|\rho_{nm} \mbox{Tr}_{\rm E}\left[\mathcal{B}_n(\tau) \rho_{\rm pn} \mathcal{B}_{m}^\dagger(\tau)\right]\nonumber\\
&=& \sum_{n = 0}^N|n\rangle\langle n|\rho_{nn} + \sum_{n \neq m = 1}^N |n\rangle\langle m|\rho_{nm} \mbox{Tr}_{\rm E}\left[B_-^{(n)}U_{\rm P}(\tau) B_+^{(n)} \rho_{\rm pn} B_-^{(m)}U_{\rm P}^\dagger(\tau) B_+^{(m)}\right]\nonumber\\
&=& \sum_{n = 1}^N|n\rangle\langle n|\rho_{nn} + \sum_{n \neq m = 1}^N |n\rangle\langle m|\rho_{nm} \mbox{Tr}_{\rm E}\left[B_-^{(m)}B_+^{(m)}(\tau)B_-^{(n)}(\tau)B_+^{(n)}\rho_{\rm pn}\right],
\end{eqnarray}
where $|0\rangle$ is the collective ground state. Since the phonon baths are assumed to be uncorrelated, 
\begin{equation}
\rho(\tau) = \sum_{n = 0}^N|n\rangle\langle n|\rho_{nn} + \mathcal{C}^2(\tau) \sum_{n \neq m = 1}^N |n\rangle\langle m|\rho_{nm}. 
\label{system_state}
\end{equation}
Therefore for $\tau = \tau_{\rm P}$ we obtain the system state,
\begin{equation}
\rho(\tau_{\rm P}) = \sum_{n = 0}^N|n\rangle\langle n|\rho_{nn} + \mathcal{C}_\infty^2(T) \sum_{n \neq m = 1}^N |n\rangle\langle m|\rho_{nm}. 
\label{system_state2}
\end{equation}
Equation~\eqref{system_state} describes the consequence of the non-markovian linear phonon coupling that introduces the factor $\mathcal{C}^2(\tau)$ in the off-diagonal elements of the matrix, which may lead to fast picosecond scale initial decoherence. One would miss this when not accounting for the linear phonon coupling in the equations of motion. Equation~\eqref{system_state2} on the other hand serves as an input to the further evolution described by the markovian master equation after the phonon bath has been relaxed (for $\tau \geq \tau_{\rm P}$) in our concatenation scheme.
\twocolumngrid


\bibliography{references}

\end{document}